\documentclass[]{spie}

\usepackage{amsmath,amsfonts,amssymb}
\usepackage{graphicx}
\usepackage{color}
\usepackage{hyperref}

\usepackage{empheq}
\usepackage[export]{adjustbox}

\usepackage{caption}

\title{Time-Harmonic Optical Chirality in Inhomogeneous Space}

\author[a]{Philipp Gutsche}
\author[b]{Lisa V. Poulikakos}
\author[a]{Martin Hammerschmidt}
\author[a,c]{Sven Burger}
\author[a,c]{Frank~Schmidt}
\affil[a]{Zuse Institute Berlin, Takustr. 7, 14195 Berlin, Germany}
\affil[b]{Optical Materials Engineering Laboratory, ETH Zurich, Leonhardstr. 21,\newline 8092 Zurich,~Switzerland}
\affil[c]{JCMwave GmbH, Bolivarallee 22, 14050 Berlin, Germany}

\authorinfo{Further author information: (Send correspondence to P.G.)\\P.G.: E-mail: gutsche@zib.de}

\newcommand{\Eth}{\vec{\mathcal{E}}}
\newcommand{\Dth}{\vec{\mathcal{D}}}
\newcommand{\Hth}{\vec{\mathcal{H}}}
\newcommand{\Bth}{\vec{\mathcal{B}}}
\newcommand{\Jth}{\vec{\mathcal{J}}}
\newcommand{\chith}{\mathfrak{X}}
\newcommand{\chielth}{\mathfrak{X}_\text{e}}
\newcommand{\chimath}{\mathfrak{X}_\text{m}}
\newcommand{\Sigmth}{{\vec{\mathfrak{S}}}}
\newcommand{\Sth}{{\vec{\mathcal{S}}}}
\newcommand{\Uelth}{\mathcal{U}_\text{e}}
\newcommand{\Umath}{\mathcal{U}_\text{m}}
\newcommand{\Uth}{\mathcal{U}}

\newcommand{\chiTelth}{\widetilde{\mathfrak{X}}_\text{e}}
\newcommand{\chiTmath}{\widetilde{\mathfrak{X}}_\text{m}}

\newcommand{\E}{\vec{E}}
\newcommand{\D}{\vec{D}}

\newcommand{\J}{\vec{J}}

\newcommand{\chii}{\chi}

\newcommand{\kk}{\vec{k}}
\newcommand{\xx}{\vec{x}}

\newcommand{\epsi}{\varepsilon}
\newcommand{\mue}{\mu}

\newcommand{\nabl}{\vec{\nabla}}
\newcommand{\rot}[1]{\left( \nabl \times #1 \right)}

\newcommand{\real}[1]{\operatorname{Re}\left(#1\right)}
\newcommand{\imag}[1]{\operatorname{Im}\left(#1\right)}

\renewcommand{\vec}[1]{\boldsymbol{#1}}

\newcommand{\myInt}[2]{#1^{(#2)}}
\newcommand{\dd}[1]{\partial#1}

\begin{document} 
\maketitle

\noindent
This paper will be published in
Proc.\ SPIE \textbf{9756},
\textit{Photonic and Phononic Properties of Engineered Nanostructures VI},
97560X (March 14, 2016),
\textit{doi:10.1117/12.2209551}
and is made available as an electronic preprint with permission of SPIE.
Copyright 2016 Society of Photo Optical Instrumentation Engineers.
One print or electronic copy may be made for personal use only.
Systematic electronic or print reproduction and distribution, duplication of any material in this
paper for a fee or for commercial purposes, or modification of the
content of the paper are prohibited.
\newline
\phantom{x}
\href{http://dx.doi.org/10.1117/12.2209551}{\underline{\smash{http://dx.doi.org/10.1117/12.2209551}}}

\begin{abstract}
Optical chirality has been recently suggested to complement the physically
relevant conserved quantities of the well-known Maxwell's equations.
This time-even pseudoscalar is expected to provide further insight
in polarization phenomena of electrodynamics such as spectroscopy
of chiral molecules. Previously, the corresponding continuity equation
was stated for homogeneous lossless media only. We extend the underlying
theory to arbitrary setups and analyse piecewise-constant material
distributions in particular. Our implementation in a Finite Element Method
framework is applied to illustrative examples in order to introduce
this novel tool for the analysis of time-harmonic simulations of
nano-optical devices.
\end{abstract}


\section{INTRODUCTION}
\label{sec:intro}

Polarization is an important degree of freedom of light. While many natural light sources such as the sun
emit unpolarized light, both nature and technology tailor polarization of electromagnetic waves.
Applications range from animals using polarization as means of communication \cite{gagnon2015}
to advanced quantum communication protocols such as quantum-key-distribution \cite{vallone2015}.

Especially, circularly polarized light (CPL) is of major interest because many chiral structures are very
sensitive to right- and left-handed polarized light. The term chirality describes geometrically that these
structures are not super-imposable with their mirror image \cite{kelvin1904}. Chiral shapes are ubiquitous
in nature and can be found in e.g.\ biological photonic crystals within butterfly wings \cite{saba2014}. 

In order to quantify the handedness or helicity of arbitrary electromagnetic fields
Tang and Cohen \cite{tang2010} introduced the concept of optical chirality and showed that circularly polarized
plane waves are optically chiral. 
Based on the concept of Lipkin's zilch \cite{Lipkin1964}, they stated a continuity equation for optical chirality in vacuum
and motivated its physical relevance by deriving this quantity from the excitation rate of chiral molecules. Since then it has been
shown that the optical chirality density is related to the enhancement of circular dichroism (CD) \cite{garcia2013}, i.e.\ the spectroscopy
of small particles with the help of CPL. Recently, the conservation law of helicity or optical chirality has been extended
to polarizable particles surrounded by lossless media \cite{nieto2015} and absorbing media \cite{poulikakos2015}.

It is expected that e.g.\ chiral plasmonic particles enhance CD signals of chiral molecules \cite{mcPeak2014}.
This expectation is supported by recent experiments claiming to measure the chirality of optical near-fields \cite{meinzer2013}.
Additionally, there have been various
studies of the enhancement of optical chirality in the context of time-harmonic simulations \cite{schaeferling2012} as well as theory
\cite{bliokh2011,barnett2012,fernandez2015}.
These theoretical analyses have been mostly limited to homogeneous media \cite{philbin2013} or vacuum only.

In this study, we extend the well-known concept of polarization of propagating plane waves in the far-field to optical
near-fields in arbitrary setups. We show that with the help of a generalized optical chirality density, electromagnetic
energy can be decomposed into right- and left-handed parts at each spatial point \cite{bliokh2011}.
This accomplishes the concept of metamaterials, wherein effective, i.e.\ spatially averaged, constitutive material equations can be derived
from chiral far-field effects \cite{condon1937, tellegen1948, kong1975}.

In section \ref{sec:nota}, we start by stating our notation for the derivation of the theory of conservation of time-harmonic optical chirality.
In the third section, we give an overview of optical chirality, its relation to energy and derive its point-wise continuity equation
for arbitrary, including anisotropic, media.
Subsequently, we illustrate the concept of chirality conversion, i.e.\ change of polarization, with the help of simple examples.
We identify loss, anisotropy and gradients in the material parameters as generators of converted chirality.

Finally, we establish the applicability of our theory with the help of a numerical example similar to recent experiments 
which illustrates both the close connection of chirality and energy as well as the new perspectives gained with our approach.
We conclude this analysis with a summary and possible further applications of our results.

\section{NOTATION}
\label{sec:nota}

In order to introduce notation, we state briefly Maxwell's equations \cite{jackson1998} 
for the time-harmonic electric and magnetic fields $\Eth$ and $\Hth$, respectively,
as well as the electric flux density $\Dth$ and the magnetic flux density $\Bth$ at angular frequency $\omega$: 
\begin{align}
    \nabl \times \Eth &= i \omega \Bth \label{eq:Maxwell1th} \\
    \nabl \times \Hth &= -i \omega \Dth + \Jth \label{eq:Maxwell2th},
\end{align}
where $\nabl \cdot \Bth = 0$ and for charge-free space $\nabl \cdot \Dth = 0$, which we assume in the following.
The constitutive equations are $\Dth = \epsi \Eth = \epsi_0 \epsi_r \Eth$ and $\Bth = \mue \Hth = \mue_0 \mue_r \Hth$ with the parameters 
permittivity $\epsi$ and permeability $\mue$ which consist of material-dependent relative permittivity $\epsi_r$ and vacuum
permittivity $\epsi_0$ and, respectively, the permeabilities $\mue_r$ and $\mue_0$. $\Jth$ is the impressed electric current density
which is assumed to be zero throughout this study, although the results hold for $\Jth \neq 0$ as well.

The electromagnetic field energy $\Uth = \Uelth + \Umath$ consists of an electric and a magnetic part,
$\Uelth = 1/4 \Eth \cdot \Dth^*$ and $\Umath = 1/4 \Bth \cdot \Hth^*$, respectively.
Together with the Poynting vector $\Sth = 1/2 \Eth \times \Hth^*$, it is part of the central continuity equation of electromagnetics \eqref{eq:enConsth}.

The physical relevant real-valued and time-dependent fields $\vec{X}$ are reconstructed from the time-harmonic fields $\vec{\mathcal{X}}$
with the help of
\begin{equation}
    \vec{X} = \operatorname{Re} \left[ \vec{\mathcal{X}} \exp{\left(-i \omega t\right)} \right].
    \label{eq:th}
\end{equation}
In order to obtain real, time-averaged values of physical quantities from the time-harmonic computation, the following rules\cite{jackson1998} are used:
$\overline{\vec{\mathcal{X}} \cdot \vec{\mathcal{Y}}} = \frac{1}{2} \real{\vec{\mathcal{X}} \cdot \vec{\mathcal{Y}}^*}$ for scalar
products and $\overline{\vec{\mathcal{X}} \times \vec{\mathcal{Y}}} = \frac{1}{2} \real{\vec{\mathcal{X}} \times \vec{\mathcal{Y}}^*}$
for cross products, respectively.

Furthermore, we use integral values of locally defined scalar densities $\mathcal{Y}(\xx)$ and vectorial flux densities $\vec{\mathcal{Z}}(\xx)$ at
spatial positions $\xx$.
When integrated over a volume $\Omega$ or surface $\dd{\Omega}$, we write in the following
\begin{align*}
	\mathcal{Y}^{(\Omega)} &= \int_\Omega \mathcal{Y}(\xx) ~ dV \qquad \text{and} \\
	\vec{\mathcal{Z}}^{(\dd{\Omega})} &= \int_{\dd{\Omega}} \vec{\mathcal{Z}}(\xx) \cdot d\vec{f}.
\end{align*}
This work focuses on the optical chirality rather than e.g.\ geometrical definitions related to chiral shapes. That is why
we often use the short term chirality referring solely to optical chirality in this context.

\section{CONCEPT AND BASICS}
\label{sec:concept}

We start by introducing the concept of optical chirality and its relation to energy \cite{bliokh2011} and polarization with the help of a numerical example.
In section \ref{sec:cons}, we give an overview of the conservation of integrated quantities, i.e.\ optical chirality and optical chirality flux, in
a general electromagnetic scattering problem.
Finally, we derive the continuity equation of optical chirality in arbitrary space, including lossy media \cite{poulikakos2015}.

\subsection{Chirality, Energy and Polarization}
\label{sec:chEnPo}

Intuitively, optical chirality $\chith$ at the spatial position $\xx$ is proportional to the difference of left ($\Uth_\text{L}$) and right ($\Uth_\text{R}$) circularly polarized
electromagnetic field energy \cite{bliokh2011}:
\begin{empheq}[box=\fbox]{align}
	\label{eq:chEn}
	\chith(\xx) &= \frac{\omega n}{c_0} \left[ \Uth_\text{L}(\xx) - \Uth_\text{R}(\xx) \right] \\
	\label{eq:chEnEn}
	\Uth(\xx) &= \Uth_\text{L}(\xx) + \Uth_\text{R}(\xx),
\end{empheq}
where $n$ is the refractive index, $c_0$ is the speed of light in vacuum and the definition of the optical chirality $\chith$ will be discussed in more detail
in section \ref{sec:chCont}. The electromagnetic field energy $\Uth$
can be split into left and right circularly polarized parts, respectively. This decomposition is well understood in
the far-field, where polarization is defined with respect to the propagation direction. Here, we show that with the help
of optical chirality, circular polarization can be defined in the near-field as well.

We analyse a twisted photonic crystal fibre which has recently been found to exhibit enhanced optical activity \cite{wong2015}.
The cross section of such a device is depicted in Fig.\ \ref{fig:PhCFiber} (a). Along the out-of-plane propagation
direction, the fibre is twisted around its central axis yielding modes with preferred circular polarization.
The twist of the analysed setup is $5 \cdot 10^4 ~ \text{rad}/\text{m}$ and the vacuum wavelength $\lambda_0 = 1.55\mu\text{m}$ fixes
the out-of-plane wave vector $k_z$ of the propagating mode problem. The pitch of the hexagonal lattice of air holes is $p = 6.75\mu\text{m}$ and
the silica fibre with refractive index $n = 1.444$ is modelled infinitely large using Perfectly Matched Layers (PMLs) \cite{schaedle2007}.

In Fig.\ \ref{fig:PhCFiber} (b) and (c), we show the decomposition of electromagnetic field energy density into
parts of right (RPL)
and left (LPL) circularly polarized light. From \eqref{eq:chEn} and \eqref{eq:chEnEn}, it follows
\begin{align}
	\label{eq:U_R}
	\Uth_\text{R} &= \frac{1}{2} \left( \Uth - \frac{c_0}{\omega n} \chith \right) \\
	\Uth_\text{L} &= \frac{1}{2} \left( \Uth + \frac{c_0}{\omega n} \chith \right).
	\label{eq:U_L}
\end{align}
The chosen propagation mode in Fig.\ \ref{fig:PhCFiber} with $n_\text{eff} = 1.44965 + 0.0017i$ shows clearly preferred right circular polarization.
Previously, the classification of modes with respect to their polarization has been done from the near-fields directly \cite{wong2012}.
The analysis with the help of optical chirality complements these findings and generalizes concepts such as
local circular polarization \cite{naruse2015} and degree of circular polarization \cite{saba2011}
which depend on the choice of a specific propagation direction. This generalization will be introduced in the following sections.

\begin{minipage}{\linewidth}
\begin{center}
\begin{tabular}{ccc} 
\includegraphics[height=4.4cm]{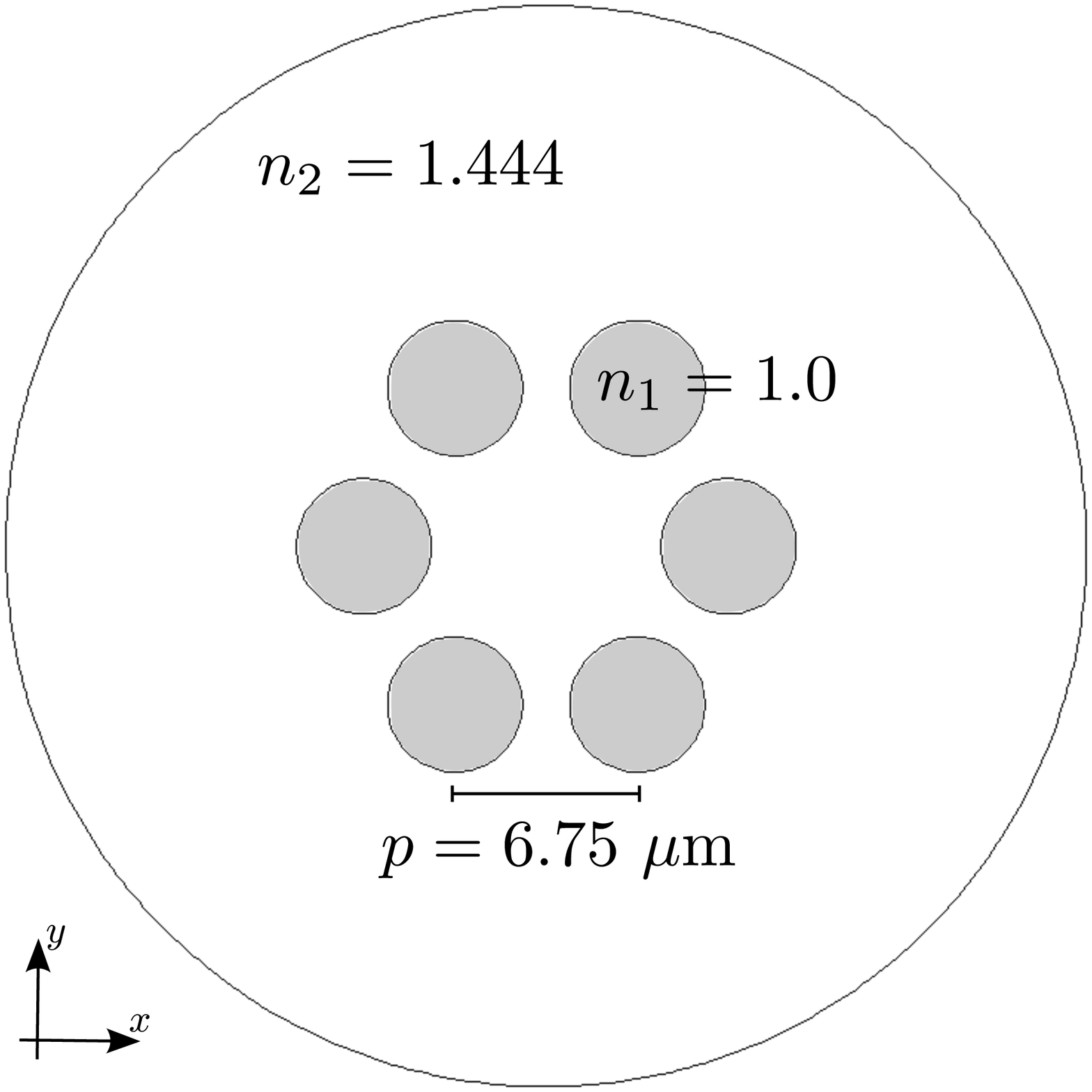} &
\includegraphics[height=5cm]{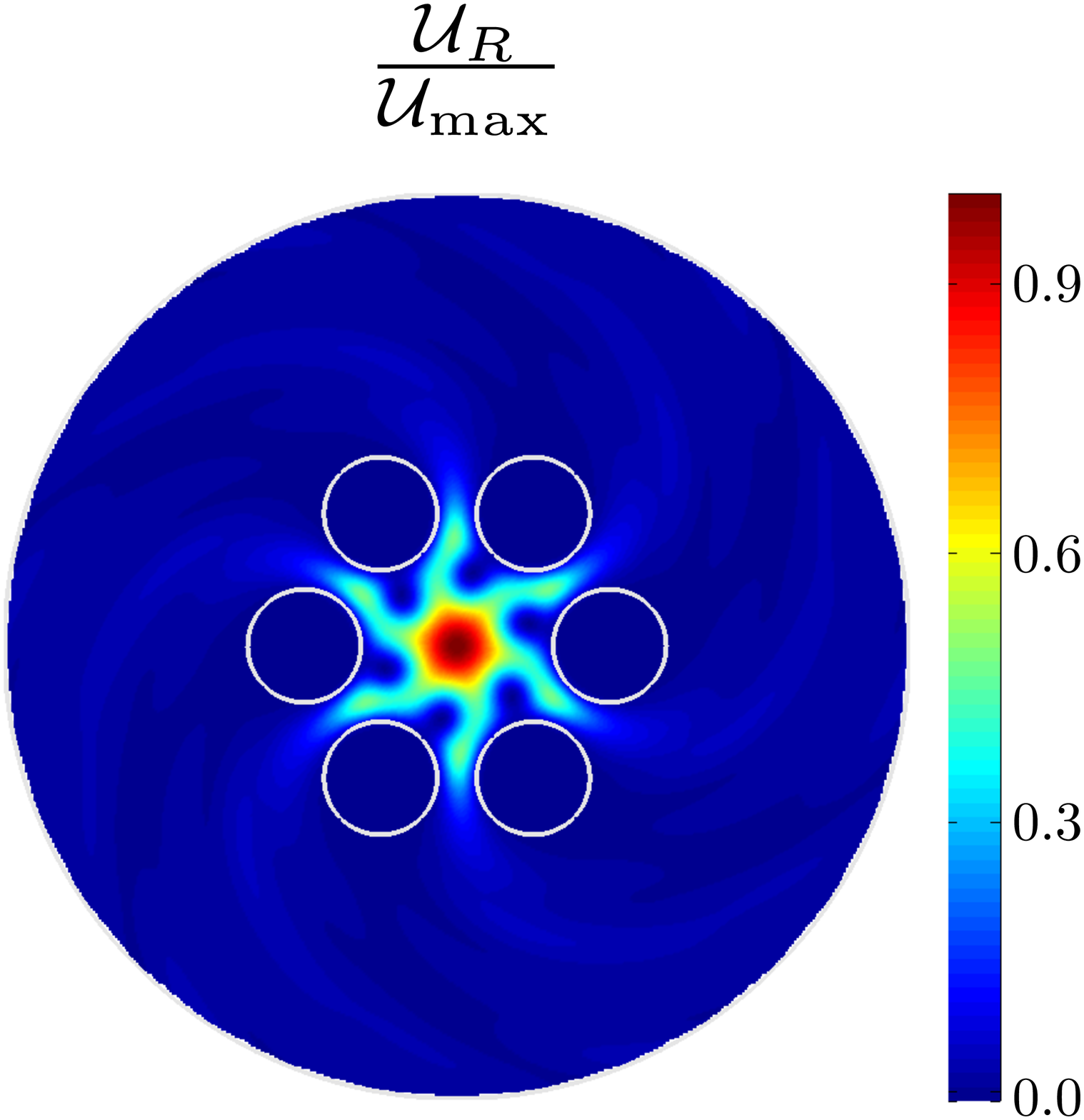} &
\includegraphics[height=5cm]{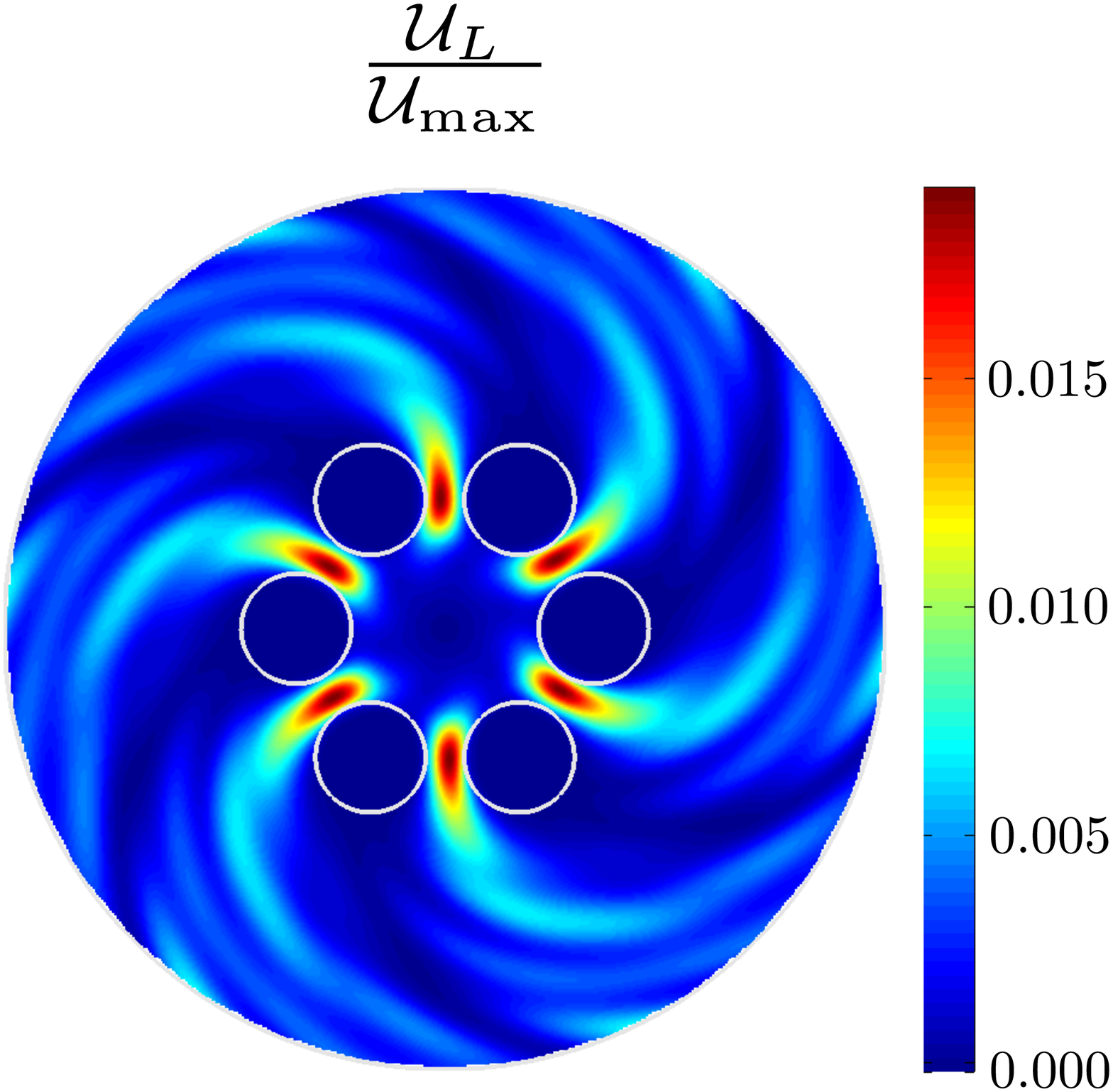} 
\\
(a) setup &
(b) right polarized energy density &
(c) left polarized energy density
\end{tabular}
\end{center}
\captionof{figure}{
	\label{fig:PhCFiber}
	Relation of optical chirality and energy of a guided mode
	in a twisted photonic crystal fibre made out of silica and air holes (a).
	The common electromagnetic field energy density is decomposed with the help of the optical chirality into right
	and left circularly polarized parts \eqref{eq:U_R}-\eqref{eq:U_L}.
	These are scaled with the maximal total field energy density $\Uth_\text{max}$.
	This specific mode shows clearly preferred right circular polarization (b) while the left circularly polarized part
	is distinctively suppressed (c).
}
\end{minipage}

\subsection{Chirality Conservation of Integrated Quantities}
\label{sec:cons}

In this section, we provide an overview of the concept of
conservation of time-harmonic optical chirality
in piecewise-constant media \cite{poulikakos2015}.
In the following sections, it will be shown that
our results are valid for any material distribution.

\begin{figure} [ht]
\begin{center}
\begin{tabular}{c} 
\includegraphics[height=7cm]{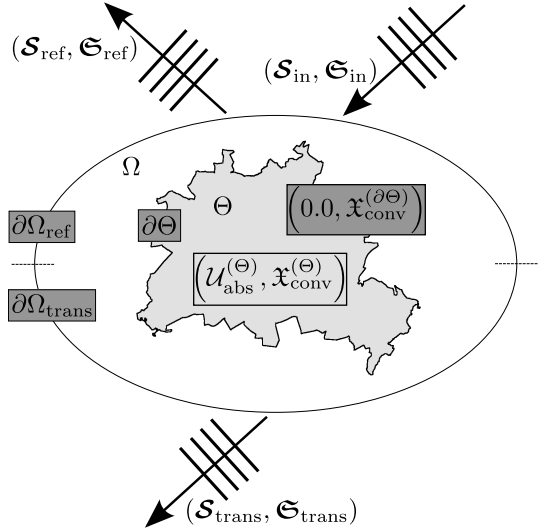}
\end{tabular}
\end{center}
\caption{
	\label{fig:setup} 
	Conceptual illustration of the conservation of energy and optical chirality for the scatterer $\Theta$ located in the surrounding $\Omega \setminus \Theta$ which is lossless for simplicity.
	The boundary $\dd{\Omega}$ is divided into upper $\dd{\Omega}_\text{ref}$ and lower $\dd{\Omega}_\text{trans}$ part. This separation is arbitrary
	for isolated but straightforward for periodic setups.
	The well-known incident energy flux $\Sth_\text{in}$ is absorbed in the scatterer ($\Uth_\text{abs}$), reflected ($\Sth_\text{ref}$) and
	transmitted ($\Sth_\text{trans}$) yielding the energy conservation \eqref{eq:enConsth}.
	Energy conservation holds for inhomogeneous media, i.e.\ there is no change of this general law introduced at the interface of the scatterer $\dd{\Theta}$
	which is indicated by the value $0.0$ on this interface.
	In the case of optical chirality, the incident chirality flux $\Sigmth_\text{in}$ is distributed into converted chirality $\chith_\text{conv}$
	in the scatterer or at its interface, the reflected chirality flux $\Sigmth_\text{ref}$	and the transmitted chirality flux $\Sigmth_\text{trans}$.
	In contrast to energy, chirality is changed at the interface of
	the scatterer ($\chith_\text{conv}^{(\dd{\Theta)}}$). The conservation law of chirality reads as \eqref{eq:chConsth}. 
}
\end{figure}

In Fig.\ \ref{fig:setup}, a general setup for scattering problems is depicted:
an incident energy flux density $\Sth_\text{in}$ is integrated over the incidence surface $\dd{\Omega}_\text{ref}$
of the computational domain $\Omega$. For source-free media ($\Jth=0$), this quantity equals the sum of
the reflected $\Sth_\text{ref}$, the transmitted energy flux $\Sth_\text{trans}$ and the absorbed energy $\Uth_\text{abs}$
in the scatterer $\Theta$:
\begin{align}
	\myInt{\Sth_\text{in}}{\dd{\Omega_\text{ref}}} = \myInt{\Sth_\text{ref}}{\dd{\Omega_\text{ref}}}
		+ \myInt{\Sth_\text{trans}}{\dd{\Omega_\text{trans}}} + \myInt{\Uth_\text{abs}}{\Theta}.
	\label{eq:enConsth}
\end{align}

In the case of optical chirality, the incoming chirality flux $\Sigmth_\text{in}$ is distributed into
reflected $\Sigmth_\text{ref}$ and transmitted $\Sigmth_\text{trans}$ chirality flux. As outlined in the following,
the scatterer's interface $\dd{\Theta}$ as well as its volume $\Theta$ might lead to a change in and conversion of chirality $\chith_\text{conv}$.
Summarizing the conservation law of optical chirality reads:
\begin{empheq}[box=\fbox]{align}
	\Sigmth_\text{in}^{(\dd{\Omega}_\text{ref})} = \Sigmth_\text{ref}^{(\dd{\Omega}_\text{ref})}
		+ \Sigmth_\text{trans}^{(\dd{\Omega}_\text{trans})} + \chith_\text{conv}^{(\Theta)} + \chith_\text{conv}^{(\dd{\Theta})}.
	\label{eq:chConsth}
\end{empheq}
The different terms of this result will be derived and motivated in detail in the following sections.

Note that this conservation law of integrated quantities can also be formulated for the scattered field, yielding:
$\myInt{\Sigmth_\text{ext}}{\dd{\Omega}} = \myInt{\Sigmth_\text{sc}}{\dd{\Omega}} + \myInt{\chith_\text{conv,tot}}{\Theta}$,
where $\Sigmth_\text{ext}$ is the extinction of optical chirality analogous to extinction of energy \cite{bohren1998}.
$\Sigmth_\text{sc}$ is the scattered optical chirality flux and $\chith_\text{conv,tot}$ is the total converted chirality,
consisting of surface and volume contributions.

Assuming that $\Omega\setminus\Theta$ is lossless and isotropic, it is sufficient to analyse the conservation law for the lossless surrounding
in order to obtain the total converted chirality $\myInt{\chith_\text{conv,tot}}{\Theta}$ of the arbitrary, including lossy, scatterer:
One computes the extinction $\myInt{\Sigmth_\text{ext}}{\dd{\Omega}}$
and scattered flux $\myInt{\Sigmth_\text{sc}}{\dd{\Omega}}$ of optical chirality on the outer boundary $\dd{\Omega}$ of the whole domain. The difference
of these two terms, i.e.\ the violation of conservation of optical chirality in a lossless medium, determines the total converted chirality
$\myInt{\chith_\text{conv,tot}}{\Theta}$.
Accordingly, the continuity equation of densities \eqref{eq:chContth} valid at all spatial points in arbitrary media is not
required for the conservation law of integrated quantities \eqref{eq:chConsth}, if only the overall influence of the scatterer on optical chirality
$\myInt{\chith_\text{conv,tot}}{\Theta} =
\chith_\text{conv}^{(\Theta)} + \chith_\text{conv}^{(\dd{\Theta})}$ is of interest.

Since optical chirality is proportional to helicity in time-harmonic formulation, the recently introduced conservation law of helicity
for lossless media \cite{nieto2015} yields an equal result as \eqref{eq:chConsth} for the total converted chirality $\myInt{\chith_\text{conv,tot}}{\Theta}$
of a polarizable particle embedded in a non-absorbing surrounding.
Nieto-Vesperinas calls the converted chirality \textit{absorbed} helicity\cite{nieto2015}.
However, we prefer the term \textit{conversion}, since optical chirality can be both generated and annihilated. Furthermore, the
formalism outlined in the following introduces the extension of the spatially dependent optical chirality density $\chith(\xx)$
to arbitrary, including lossy, media \cite{poulikakos2015}.
This enables the further investigation of the conservation of optical chirality, e.g.\ the distinction of surface and
volume contributions of chirality conversion.

\subsection{Chirality Continuity Equation of Spatially Dependent Densities}
\label{sec:chCont}

In the case of conservation of energy, the standard derivation of the time-harmonic continuity equation of energy
starts with the rate at which the electromagnetic fields do work on a charged object $\Jth^* \cdot \Eth$, where $\Jth$ is
the free current density \cite{jackson1998}. With the help of \eqref{eq:vecId}, it follows
\begin{align}
	2 i \omega &(\Uelth - \Umath) + \nabl \cdot \Sth = - \frac{1}{2} \Jth^* \cdot \Eth
	\label{eq:poyntth}
\end{align}
for homogeneous as well as inhomogeneous space. The real part of this equation represents conservation of energy.

For the derivation of the continuity equation of the time-dependent optical chirality density, it was suggested
to study the time-dependent quantity $[\J \cdot \rot{\E} + \E \cdot \rot{\J}]$ \cite{tang2010}.
The corresponding time-harmonic optical chirality density has been derived by using the well-known relation between
physically relevant and time-harmonic quantities \eqref{eq:th}\cite{schaeferling2012}.

However, we use a different approach and start with the time-harmonic quantity $[\Jth^* \cdot \rot{\Eth} + \Eth \cdot \rot{\Jth^*}]$ \cite{poulikakos2015}.
By using this starting point, we neither require homogeneity nor isotropy to obtain the continuity equation of
time-harmonic optical chirality (App. \ref{app:cont})
\begin{empheq}[box=\fbox]{align}
	2 i \omega (\chielth - \chimath) + \nabl \cdot \Sigmth
		= -\frac{1}{4} \left[ \Jth^* \cdot \rot{\Eth} + \Eth \cdot \rot{\Jth^*} \right].
	\label{eq:chContth}
\end{empheq}
Here, we distinguish between \textit{electric chirality} density $\chielth$
and \textit{magnetic chirality} density $\chimath$ for arbitrary media in analogy to electric and magnetic energy density
in \eqref{eq:poyntth}. These quantities and the 
\textit{electromagnetic chirality flux} density $\Sigmth$ are defined as\cite{poulikakos2015}
\begin{align}
    \chielth &= \frac{1}{8} \left[ \Dth^* \cdot \rot{\Eth} + \Eth \cdot \rot{\Dth^*} \right] \label{eq:chielth} \\
    \chimath &= \frac{1}{8} \left[ \Hth^* \cdot \rot{\Bth} + \Bth \cdot \rot{\Hth^*} \right] \label{eq:chimath} \\
    \Sigmth &= \frac{1}{4} \left[ \Eth \times \rot{\Hth^*} - \Hth^* \times \rot{\Eth} \right]. \label{eq:sigmth}
\end{align}
The total optical chirality is $\chith = \chielth + \chimath$.
The corresponding time-dependent chirality density $\chii = \chii_\text{e} + \chii_\text{m}$ following from these time-harmonic
definitions is similar to the literature for homogeneous media \cite{philbin2013}.
It has units of a force density and both the densities and the flux are constant
for circularly polarized propagating plane waves, i.e.\ the latter are eigenstates of optical chirality.
Since for CPL these chiral quantities differ in sign,
they enable us to distinguish between RPL and LPL in the far- and near-field.

Furthermore, they vanish for linearly polarized plane waves and have accordingly a well defined interpretation
in the far-field, where it exists a clear propagation direction given by the wave vector $\kk$.
The presented continuity equation \eqref{eq:chContth}, however, does not require the specification of any propagation direction
and is purely based on the near-field. Furthermore, there is no limitation to neither homogeneous nor
isotropic media.

It was stated that Lipkin's zilch, or optical chirality, are higher-order versions of helicity \cite{bliokh2013}.
Nevertheless, the above given quantities are easily accessible in common
time-harmonic numerical simulations and provide valuable tools for analysing
the behaviour of a wide range of nanophotonic setups in e.g.\ finite element method\cite{monk2003} (FEM) computations.
This is because of the fact that the helicity quantities derived from the suggested vector potentials of
the dual electromagnetism \cite{bliokh2013} involve rotations of these potentials.
This means the definition of helicity using dual vector potentials is not as easily accessible as the proposed optical chirality
based on $\Eth, \Dth, \Hth$ and $\Bth$.

\section{CHIRALITY CONVERSION}

While energy conservation predicts a change of the total energy only in lossy media, chirality might be changed by various
mechanisms. We call this variation the \textit{conversion of chirality} or \textit{converted chirality}.
From the formalism above, this quantity follows as
\begin{empheq}[box=\fbox]{align*}
	\chith_\text{conv} = -2 \omega \imag{\chielth - \chimath}
\end{empheq}
comparable to absorbed energy $\Uth_\text{abs} = - 2 \omega \imag{\Uelth - \Umath}$.
In the following, we analyse three kinds of material distributions yielding chirality conversion with simple examples.
For simplicity we analyse mostly non-magnetic media ($\mue_r = 1$), although our results are valid directly for $\mue \neq 0$ as well.
Loss, anisotropy and a gradient of the real part of the permittivity $\epsi$ might introduce a non-vanishing
imaginary part of $\chith$ in \eqref{eq:chielth}, i.e.\ converted chirality.

\subsection{Chirality Conversion in Isotropic Media}
\label{sec:isoMedia}

In order to discuss the novel quantities, we state the general
chirality densities (\ref{eq:chielth}, \ref{eq:chimath})
for \textbf{isotropic} homogeneous media, i.e.\ $\epsi, \mue \in \mathbb{C}$, and name them as
\begin{align}
    \chiTelth &= \frac{1}{8} i \omega \left[ \Dth^* \cdot \Bth - (\epsi \Bth)^* \cdot \Eth \right] \label{eq:chielJCM} \\
    \chiTmath &=  \frac{1}{8} i \omega \left[ \Dth^* \cdot \Bth - \Hth^* \cdot (\mue \Dth) \right] \label{eq:chimaJCM}.
\end{align}
Two important findings can be deduced from this. First, for isotropic \textbf{lossless} media, i.e.\ $\epsi, \mue \in \mathbb{R}$,
the averaged time-dependent chirality $\overline{\chii}$ is the herein introduced time-harmonic chirality as expected:
\begin{align}
	\overline{\chii} = \chith = -\frac{1}{2} \omega \imag{\Dth^* \cdot \Bth}.
	\label{eq:chIsoLossless}
\end{align}
This is in accordance to previous definitions for solely free space
of time-harmonic optical chirality
\linebreak
$\mathcal{C} = -\epsi_0 \omega/2 \imag{\Eth^* \cdot \Bth}$ \cite{schaeferling2012}.
Additionally, one can show that the magnitude of optical chirality is bounded \cite{bliokh2011}.
In our notation this yields (App. \ref{app:bound})
\begin{align*}
	\left| \chith(\xx) \right| \leq \frac{\omega n}{c_0} \Uth(\xx),
\end{align*}
which is easily understood with the help of \eqref{eq:chEn}: chirality takes its extremal value if the energy is purely
circularly polarized.

Second, the time-averaged chirality is real-valued for lossless media. However, for electrical absorbing [$\imag{\epsi} > 0$]
and magnetic absorbing [$\imag{\mue} > 0$] media,
a non-vanishing imaginary part of $\chith$ might exist.
This introduces
conversion of chirality in volumes $\chith_\text{conv}^{(\Theta)}$ in lossy media in \eqref{eq:chConsth} and further stresses the close relation between optical
chirality and electromagnetic energy \cite{bliokh2011} in homogeneous isotropic space.

In conclusion, a change of chirality in isotropic homogeneous media can only occur for lossy media. The case of piecewise constant media will be dealt with
in section \ref{sec:chConvInt}. In the next section, we analyse anisotropic media. 

\begin{figure} [b]
\begin{center}
\begin{tabular}{ccc} 
\includegraphics[height=6cm]{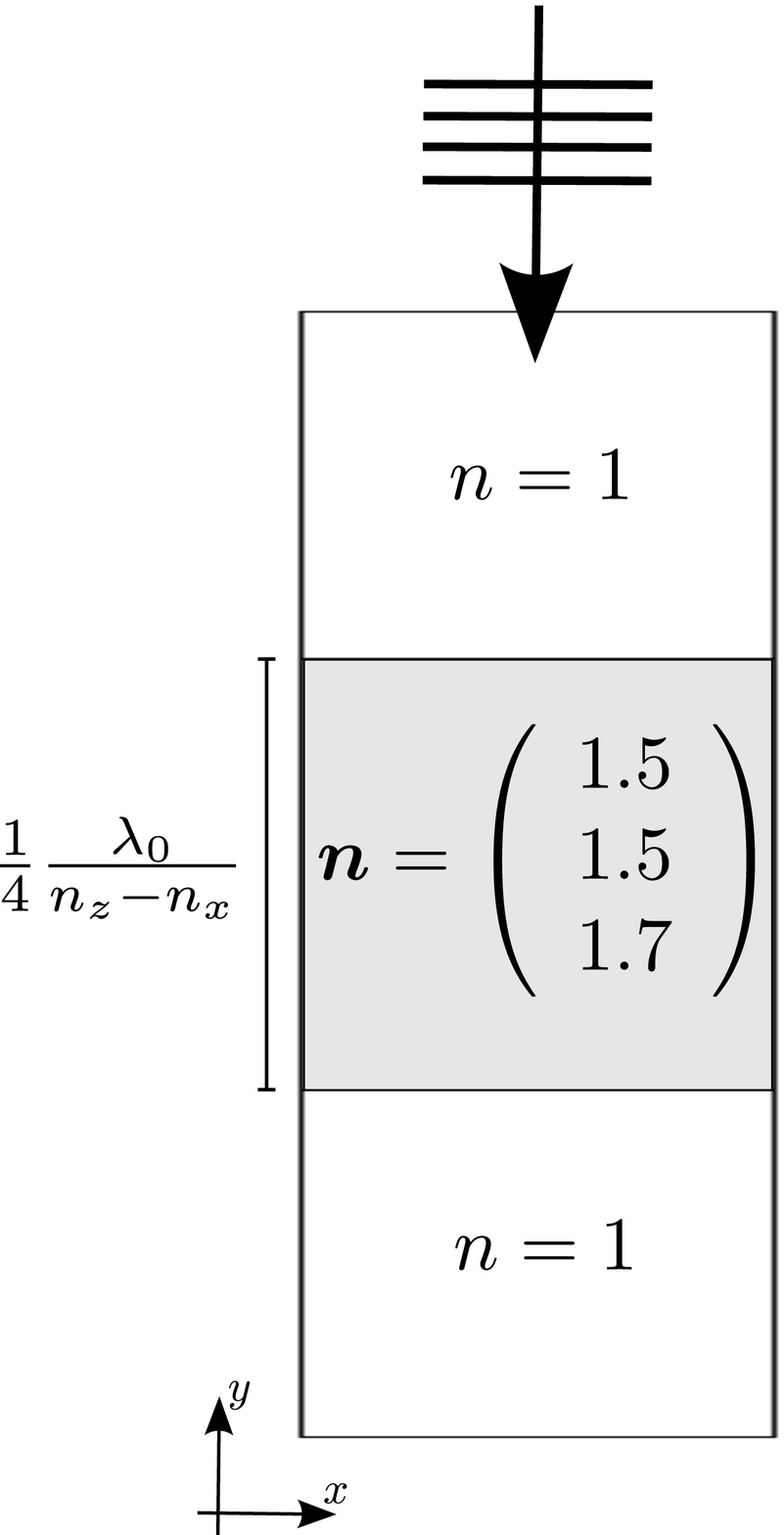} &
\hspace{1cm}
\includegraphics[height=6cm]{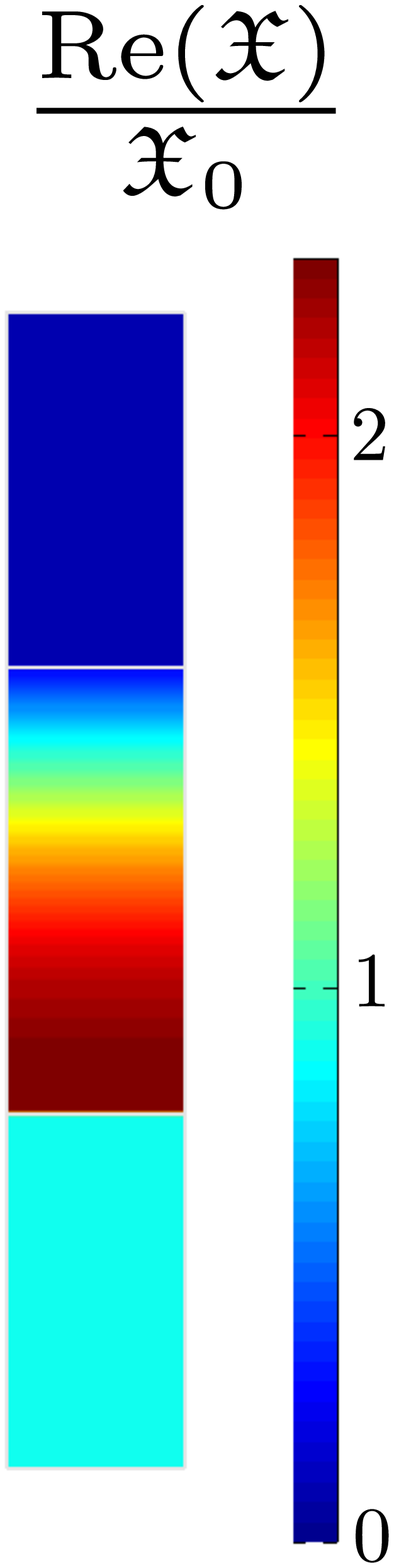} &
\hspace{1cm}
\includegraphics[height=6cm]{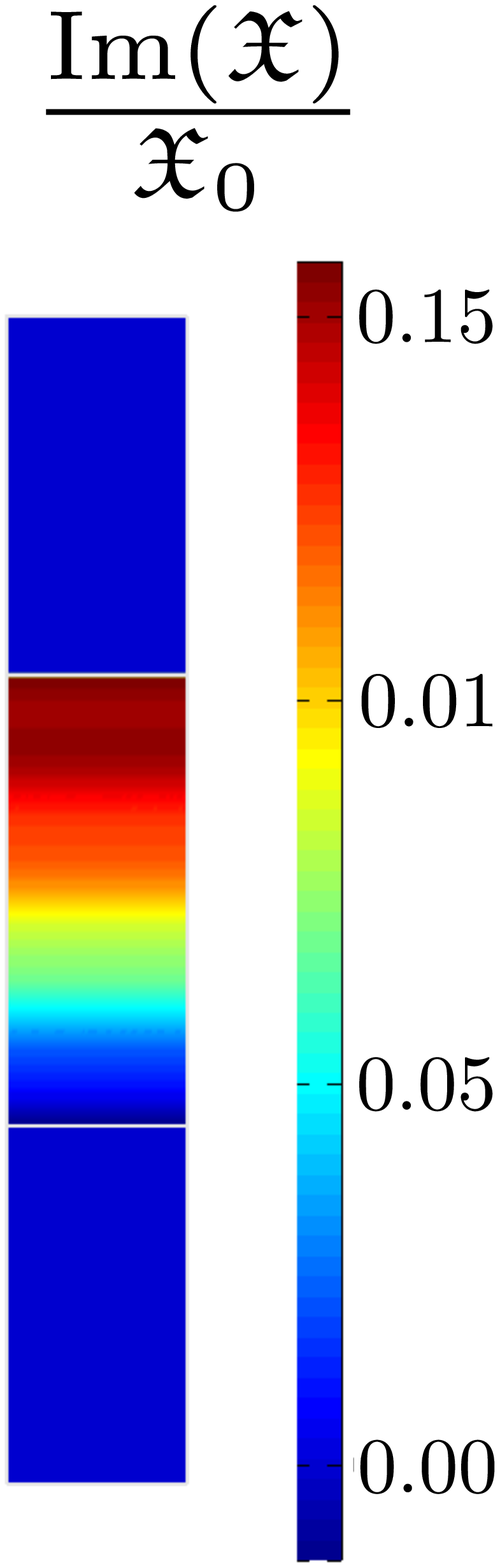} 
\\
(a) setup &
\hspace{1cm} 
(b) optical chirality &
\hspace{1cm} 
(c) converted chirality
\end{tabular}
\end{center}
\caption{
	\label{fig:quarterWave}
	Chirality conversion in anisotropic media: a quarter-wave plate with anisotropic refractive index $\vec{n}$ is illuminated from above (a). The incident plane wave
	is linearly polarized with a polarization angle of $45^\circ$ and exhibits optical chirality $\chith = 0$.
	Circularly polarized light exhibiting optical chirality $\chith \neq 0$ is generated through the anisotropy (b).
	The spatially resolved converted chirality $\chith_\text{conv} \propto \imag{\chith}$ is visible in (c).
	Note that there is no conversion in isotropic lossless media surrounding the slab, i.e.\ $\imag{\chith} = 0$ in the sub- and super-space.
	The chirality densities are scaled with the magnitude of CPL in vacuum $\chith_0$.
}
\end{figure}

\subsection{Chirality Conversion in Anistropic Media}

In the previous section, we analysed homogeneous isotropic media for which the electric and magnetic chiralities are \eqref{eq:chielJCM} and \eqref{eq:chimaJCM}.
For general material parameters $\epsi$ and $\mue$, these quantities are given by \eqref{eq:chielth} and \eqref{eq:chimath}.
Since in the previous case we identified lossy media as being capable of changing chirality, we restrict the analysis to \textbf{lossless anisotropic} media
[$\imag{\epsi} = \imag{\mue} = 0$], here. 

Due to the close connection of optical chirality and circular polarization (Sec. \ref{sec:chEnPo}), we analyse a quarter-wave plate.
It is well-known that
this simple device is used to convert linear into circular polarization \cite{hecht2002}.
In Fig.\ \ref{fig:quarterWave} (c) the conversion of chirality proportional to
$\operatorname{Im}[\chith(\xx)]$ in the anisotropic slab is clearly visible.
Additionally, the out-going plane wave carries optical chirality $\operatorname{Re}[\chith(\xx)]$, i.e.\ it possesses elliptical polarization.

Not only the anisotropy of the slab, but also the discontinuities at the air-plate and plate-air interface introduce
changes, i.e.\ conversion, of chirality.
This interface contribution $\chith_\text{conv}^{(\dd{\Theta})}$ will be further analysed in the next section.

\subsection{Chirality Conversion on Interfaces}
\label{sec:chConvInt}

As derived in the previous two sections conversion of chirality, i.e.\ $\imag{\chith} \neq 0$, can either be achieved
by loss or anisotropy in homogeneous space.
Here, we show that conversion of chirality also occurs for spatially dependent material parameters.
We restrict the analysis to isotropic space and non-magnetic media ($\mue_r = 1$), here.

Generalizing the results of section \ref{sec:isoMedia} to spatially varying $\epsi$, we obtain for the imaginary part
of the electric chirality (App. \ref{app:convInt})
\begin{align}
	8 \imag{\chielth} = 2 \omega \imag{\epsi} \imag{\Eth^* \cdot \Bth} +
		\left[ \nabl \real{\epsi} \right] \cdot \imag{\Eth^* \times \Eth}.
	\label{eq:chConvInt}
\end{align}
Similar results can be obtained for the magnetic chirality with $\mue_r \neq 1$.
The first term corresponds to the findings in section \ref{sec:isoMedia}. The second term shows that a gradient of the real part of the
permittivity yields a change in chirality.
Again, from the close connection of chirality and polarization this fact is well-known: specially designed prisms are used
to convert linear into circular polarization similar to the previously analysed quarter-wave plate.

\begin{figure} [b]
\begin{center}
\begin{tabular}{ccc} 
\includegraphics[height=4.5cm]{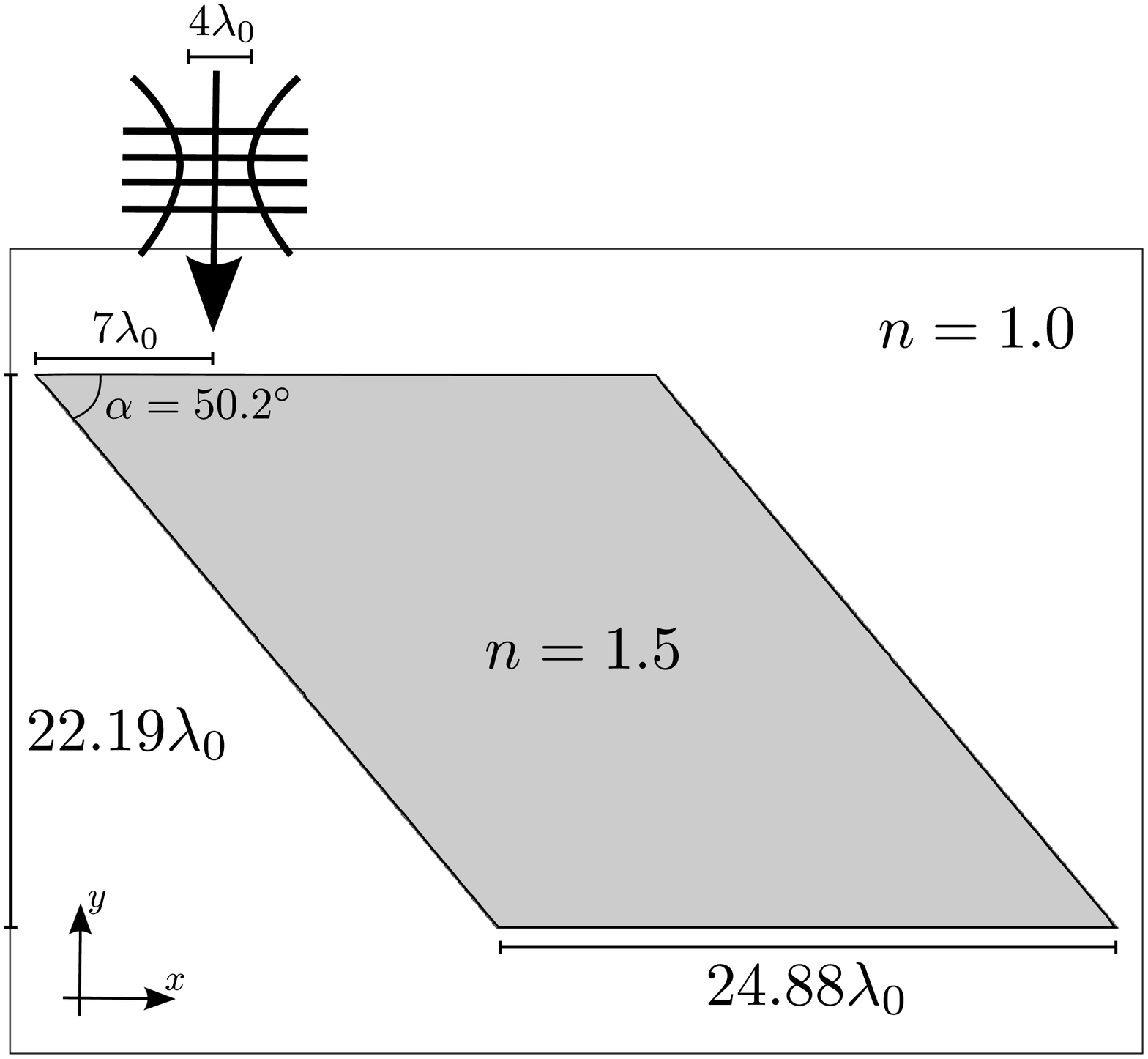} &
\includegraphics[height=4.5cm]{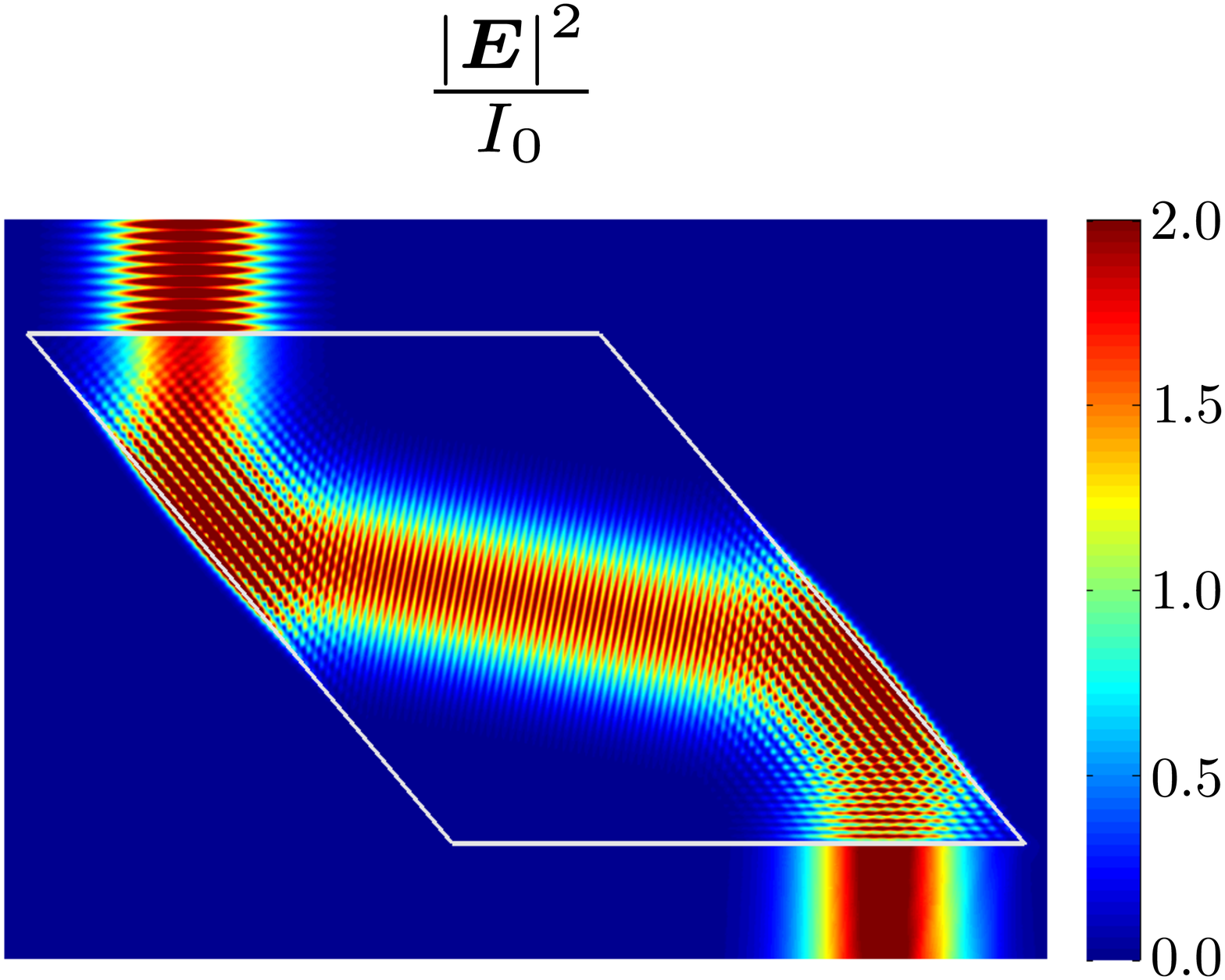} &
\includegraphics[height=4.5cm]{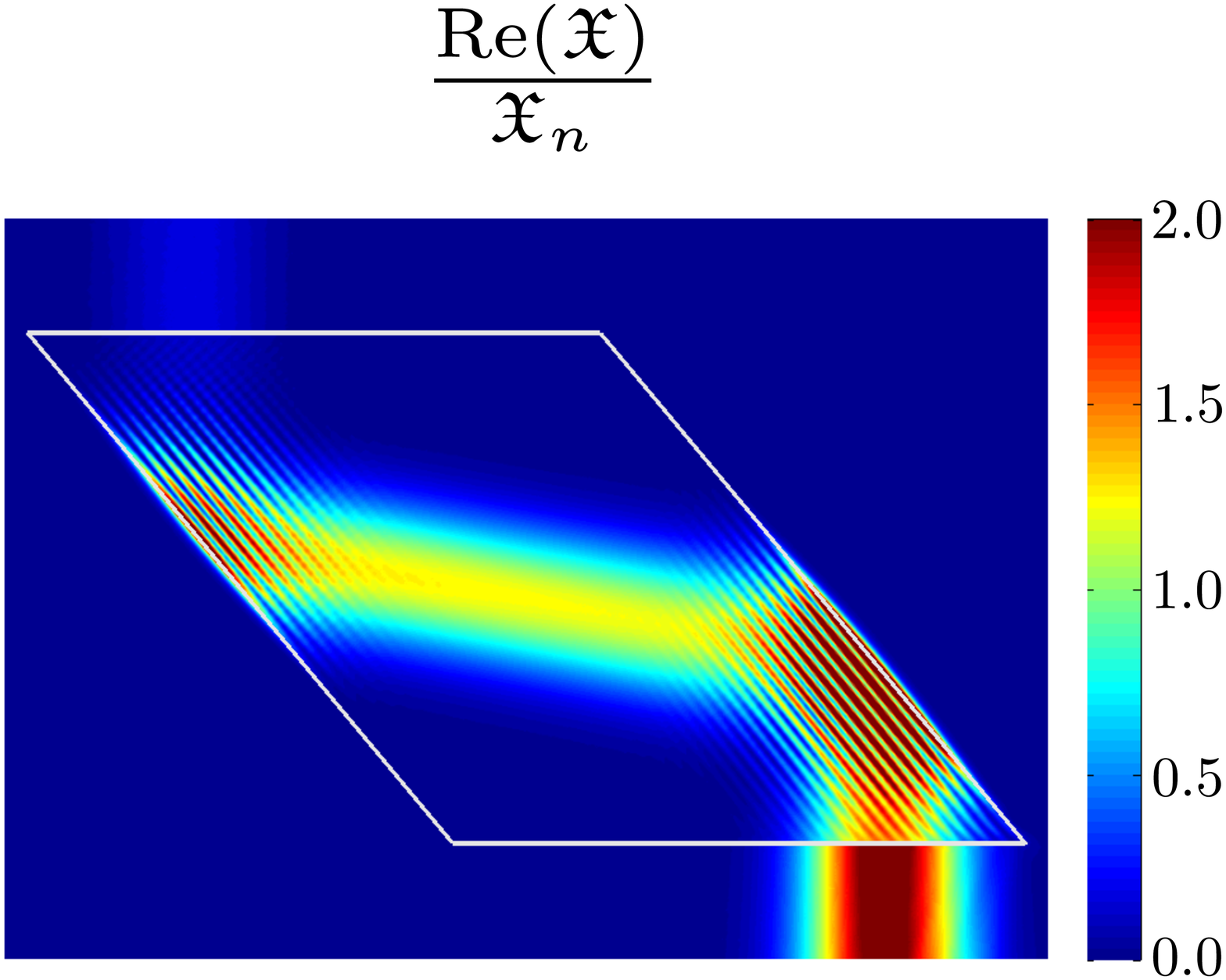} 
\\
(a) setup &
(b) intensity &
(c) optical chirality
\end{tabular}
\end{center}
\caption{
	\label{fig:fresnel}
	Chirality conversion on lossless interfaces:
	Fresnel rhomb with refractive index $n = 1.5$ and angle $\alpha$ chosen in such a way that both total internal reflection and a
	$\pi / 4$ phase shift
	between $x$- and $z$-components is introduced (a). The incident beam is approximated by a superposition of plane waves and its
	polarization angle is $45^\circ$. The dimensions of the rhomb support constructive interference along the beam path.
	The resulting intensity distribution (b) is scaled with the vacuum intensity $I_0$ of the sum of the plane waves.
	The appropriate angle $\alpha$ yields elliptically polarized light at the first total internal reflection (c).
	Together with the second reflection, CPL is generated at the lower output of the rhomb which is indicated by
	non-vanishing optical chirality $\real{\chith}$. The field is scaled with the plane wave chirality
	$\chith_n(\xx) = 0.5 I_0 \omega \epsi_0/c_0 n(\xx)^3$ depending on the refractive index at the spatial position $\xx$.
}
\end{figure}

In a Fresnel rhomb, linearly polarized light is twice reflected under an angle larger than the critical angle 
of total internal reflection \cite{fresnel1832}. Choosing the angle of this prism in such a way that a $45^\circ$ phase change between
$s$- and $p$ polarization is introduced at each reflection yields the transformation of linearly polarized to circularly polarized light. The spatial variation of
the according optical chirality is depicted in Fig.\ \ref{fig:fresnel} (c) for an incident beam.

For piecewise constant $\epsi$, the gradient of the permittivity vanishes everywhere except at domain interfaces where it
is ill-defined due to a discontinuity of $\epsi$. Nevertheless, we can compute the chirality density \eqref{eq:chielth} for each
domain with constant $\epsi$. Here, the change of conservation of the integrated optical chirality occurs only at interfaces
and is therefore associated with a conversion of chirality $\myInt{\chith_\text{conv}}{I}$ due to the interface $I$. 

Since chirality conservation for homogeneous media is valid separately in domain $D_k$ and domain $D_j$, which share the interface $I$,
the converted chirality at the interface is
\begin{align}
    \myInt{\chith_\text{conv}}{I} = \int_{\dd{D_k}} \Sigmth + \int_{\dd{D_j}} \Sigmth,
	\label{eq:absSigm}
\end{align}
where the interfaces are $I = \dd{D_k} = \dd{D_j}$ and the plus sign is due to a change of the surface normal of
$D_k$ and $D_j$, respectively.
The integration of the chirality flux \eqref{eq:sigmth}
from domain $D_k$ to $D_j$ [first summand in \eqref{eq:absSigm}] and from domain $D_j$ to $D_k$ [second summand in \eqref{eq:absSigm}]
yield different results due to the unconserved homogeneous chirality densities (\ref{eq:chielJCM}, \ref{eq:chimaJCM}).

\begin{figure}[b]
\begin{center}
\begin{tabular}{ccc} 
\includegraphics[height=5cm]{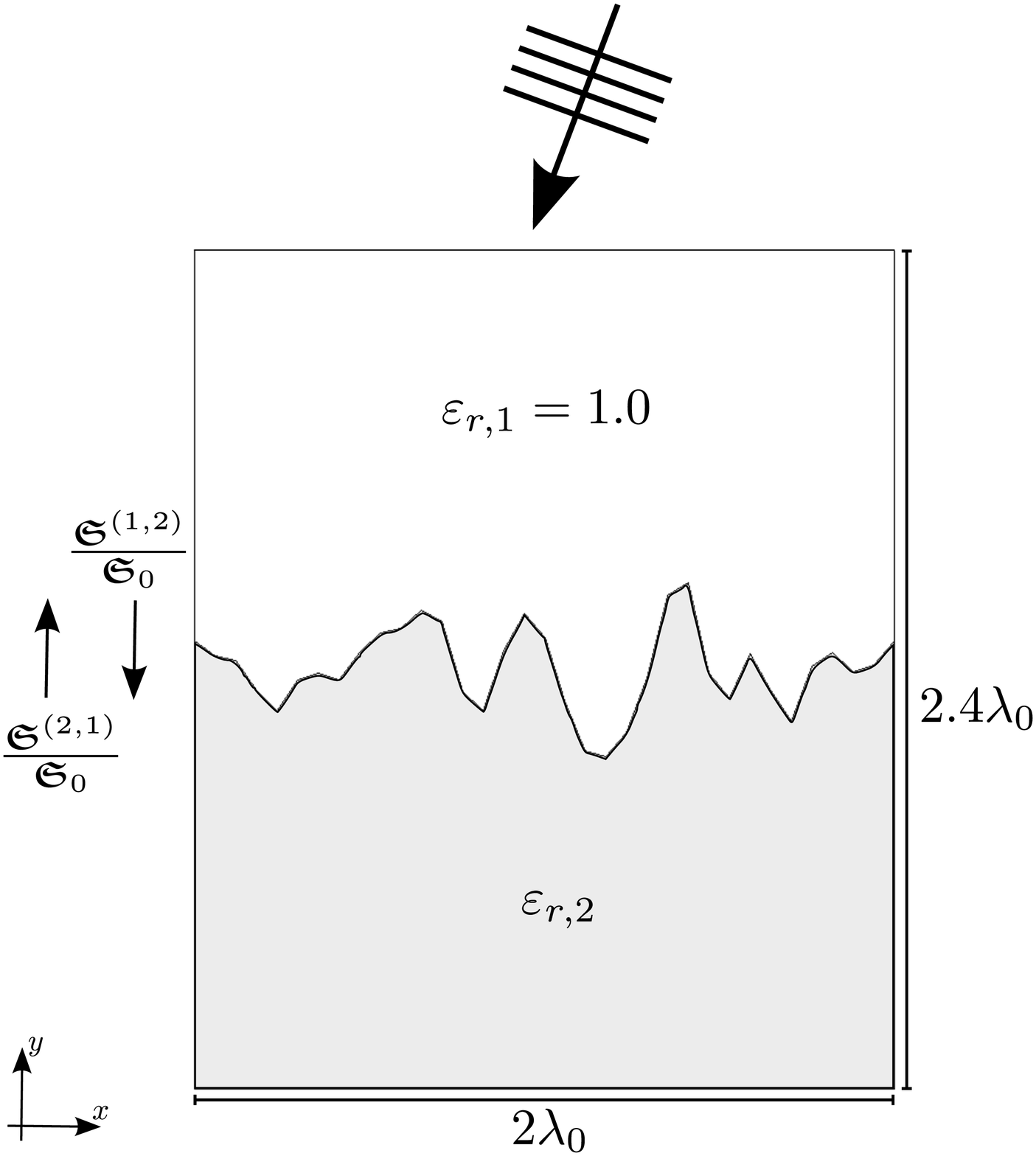} &
\includegraphics[height=5cm]{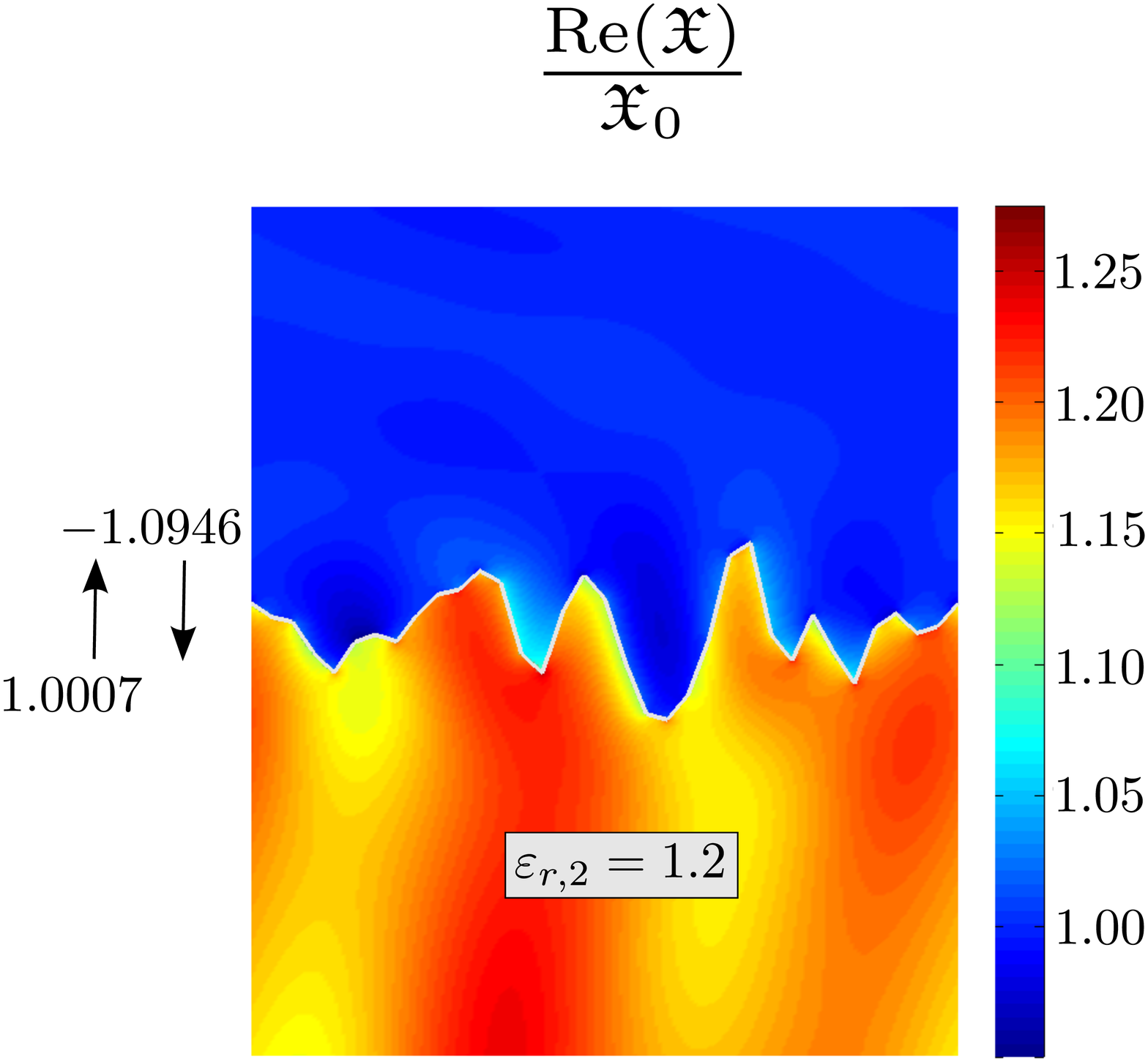} &
\includegraphics[height=5cm]{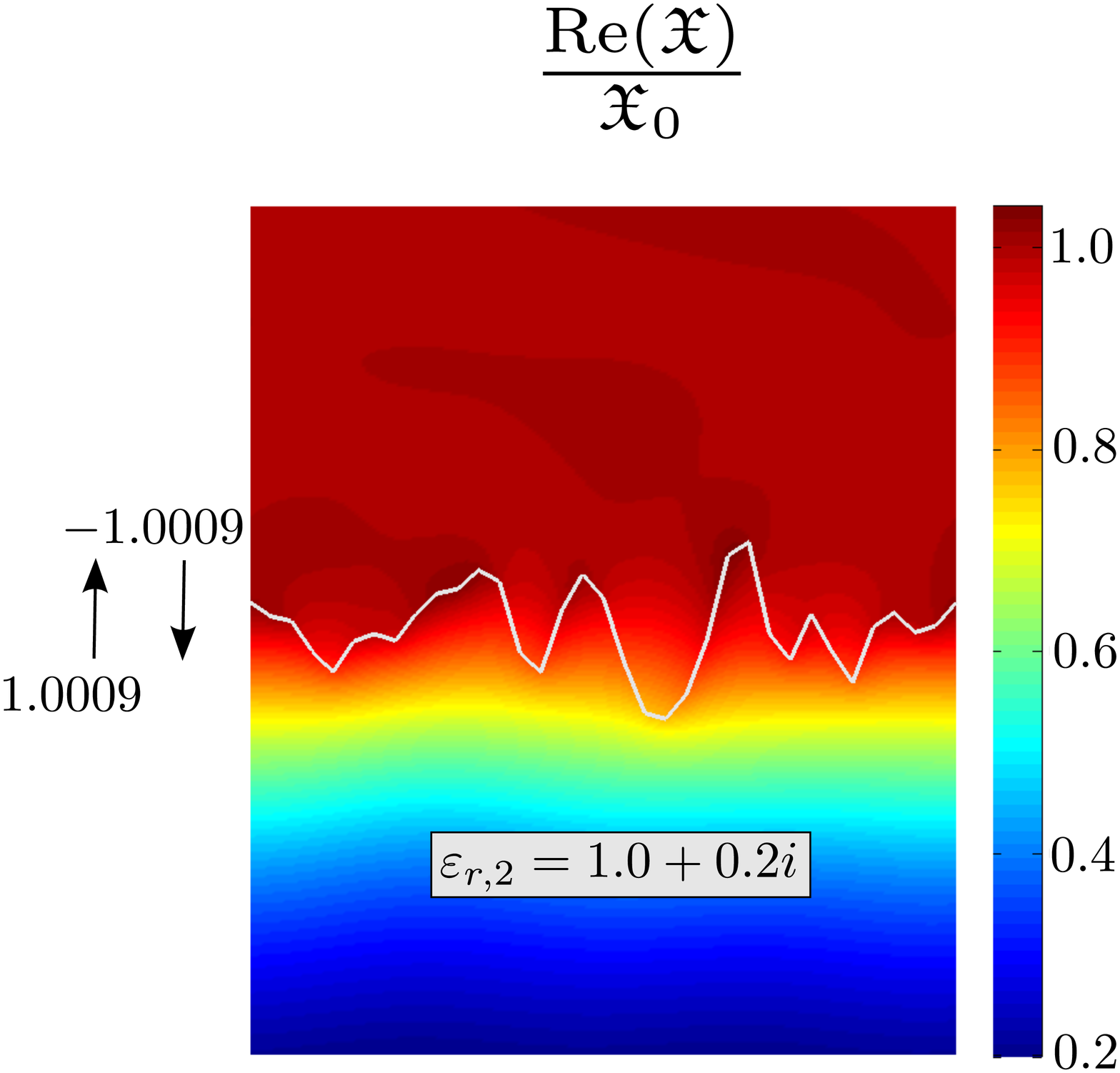} 
\\
(a) setup &
(b) lossless medium &
(c) lossy medium
\end{tabular}
\end{center}
\caption{
	\label{fig:randInt}
	Chirality conversion through a discontinuity of the permittivity $\epsi$:
	a random interface is illuminated from above with CPL rotated by $20^\circ$ around the $z$-axis (a). The arrows on the left-hand side
	indicate the two terms for the conversion of chirality at material parameter discontinuities \eqref{eq:absSigm}: $\Sigmth^{(i,j)}$ is
	the chirality flux from domain $i$ to domain $j$ and $\Sigmth_0$ is the chirality flux in vacuum over the specific interface.
	The substrate material is either lossless (b) or lossy (c). Due to the gradient in the real part of the relative permittivity $\epsi_{r,2}$,
	the lossless case introduces chirality conversion [cf.\ \eqref{eq:chConvInt}]. On the other hand, a change solely in the imaginary part of $\epsi$, i.e.\ the lossy
	case, does not convert chirality on the interface. Accordingly, the fluxes from one interface to another and vice versa are equal.
}
\end{figure}

This surface attributed conversion of chirality is shown in Fig.\ \ref{fig:randInt}. A change in the real part of the permittivity
introduces the non-vanishing term $\myInt{\chith_\text{conv}}{I}$ \eqref{eq:absSigm}
at the randomly shaped interface $I$. However, if only the
imaginary part of $\epsi$ varies over the interface, the optical chirality flux is equal in domain $D_1$ and $D_2$ yielding
$\myInt{\chith_\text{conv}}{I} = 0$. Within $D_2$, $\chith_\text{conv}(\xx) \neq 0$ as discussed in section \ref{sec:isoMedia}.

Note that if the interface is dual symmetric, i.e.\ the ratio $\epsi / \mue$ is constant for both adjacent domains, there is no chirality conversion \cite{fernandez2013}.
This is due to the fact that the contribution to $\myInt{\chith_\text{conv}}{I}$ from $\nabl \epsi$ cancels with the one due to
$\nabl \mue$. This is related to a redistribution of energy between electric and magnetic parts by a dual symmetric interface.

\pagebreak

\section{NUMERICAL APPLICATION}
\label{sec:examp}

\begin{figure}[b]
\begin{center}
\begin{tabular}{ccc} 
\includegraphics[height=6cm]{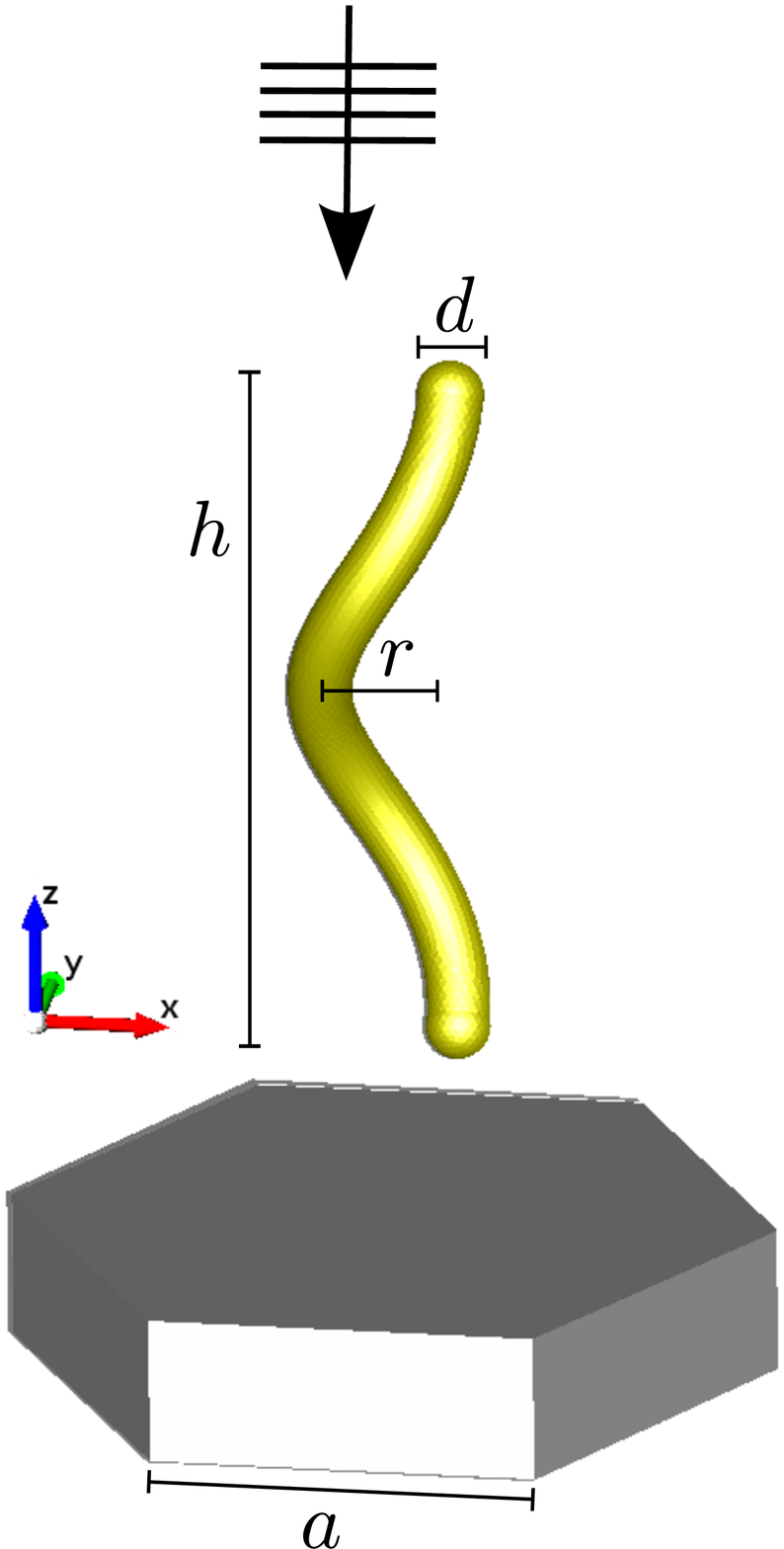}
&
\includegraphics[height=6cm]{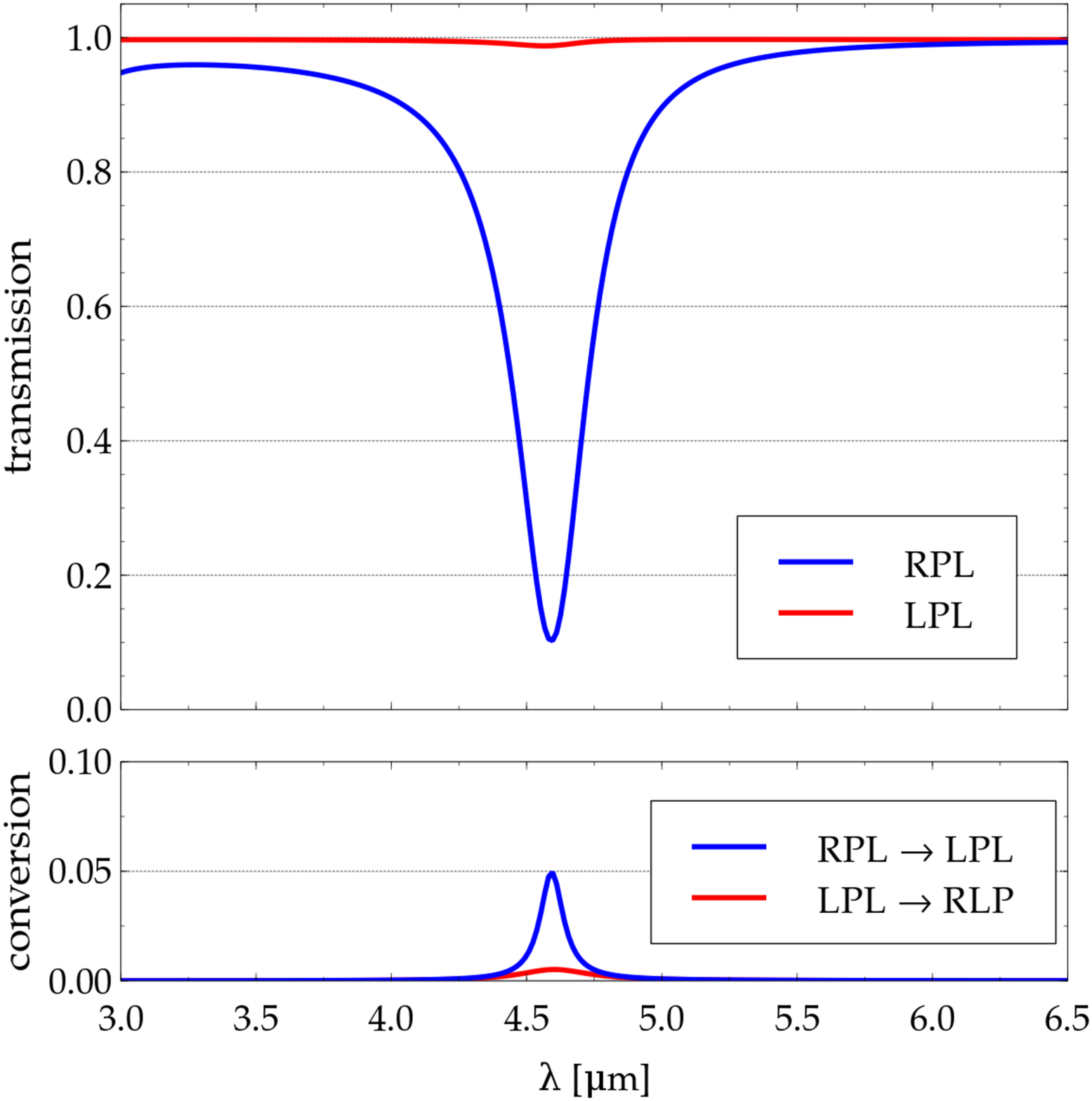}
&
\raisebox{3cm}
{
\begin{tabular}{c}
	\includegraphics[height=4.5cm]{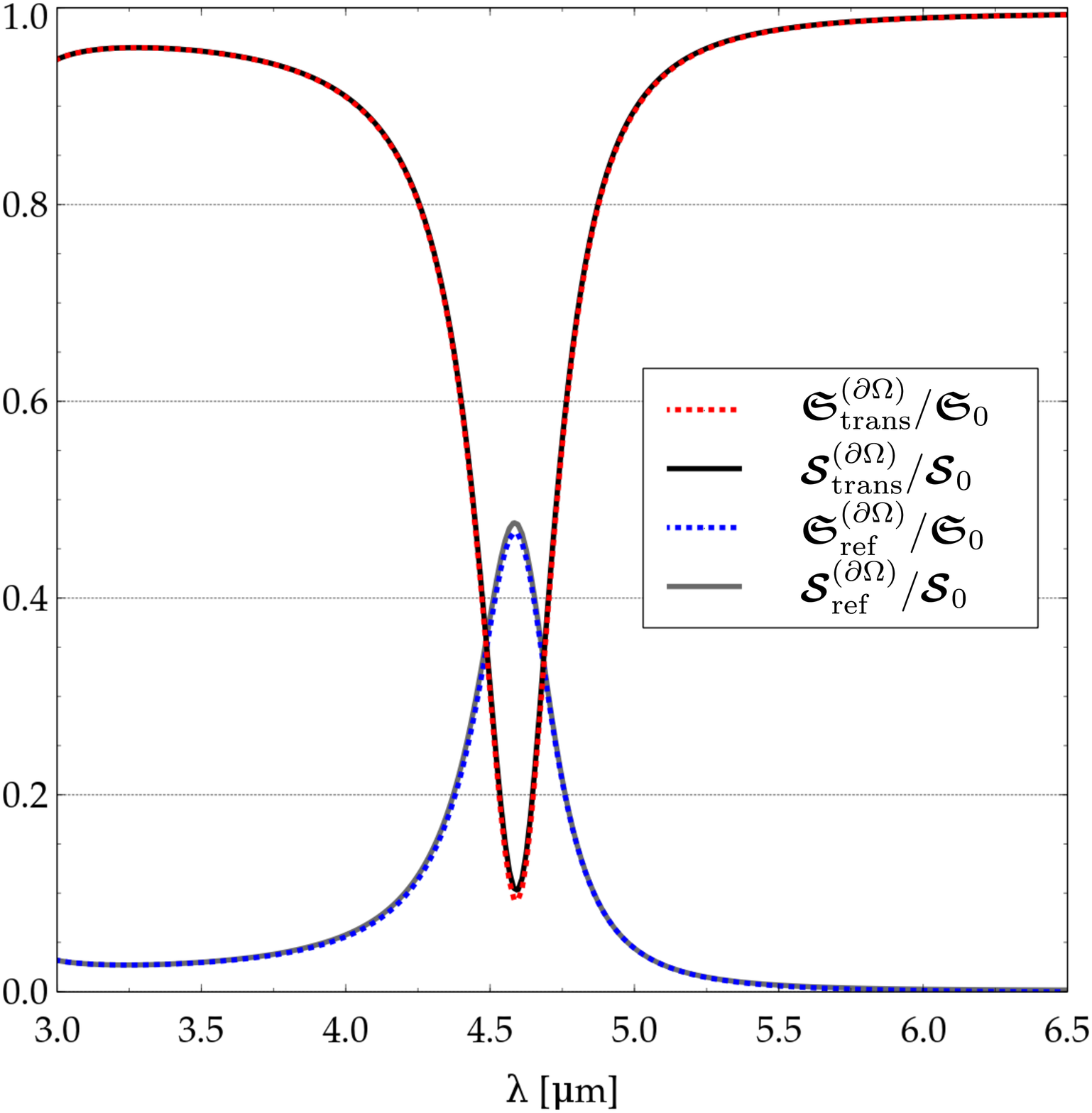} \\
	\includegraphics[height=4.5cm]{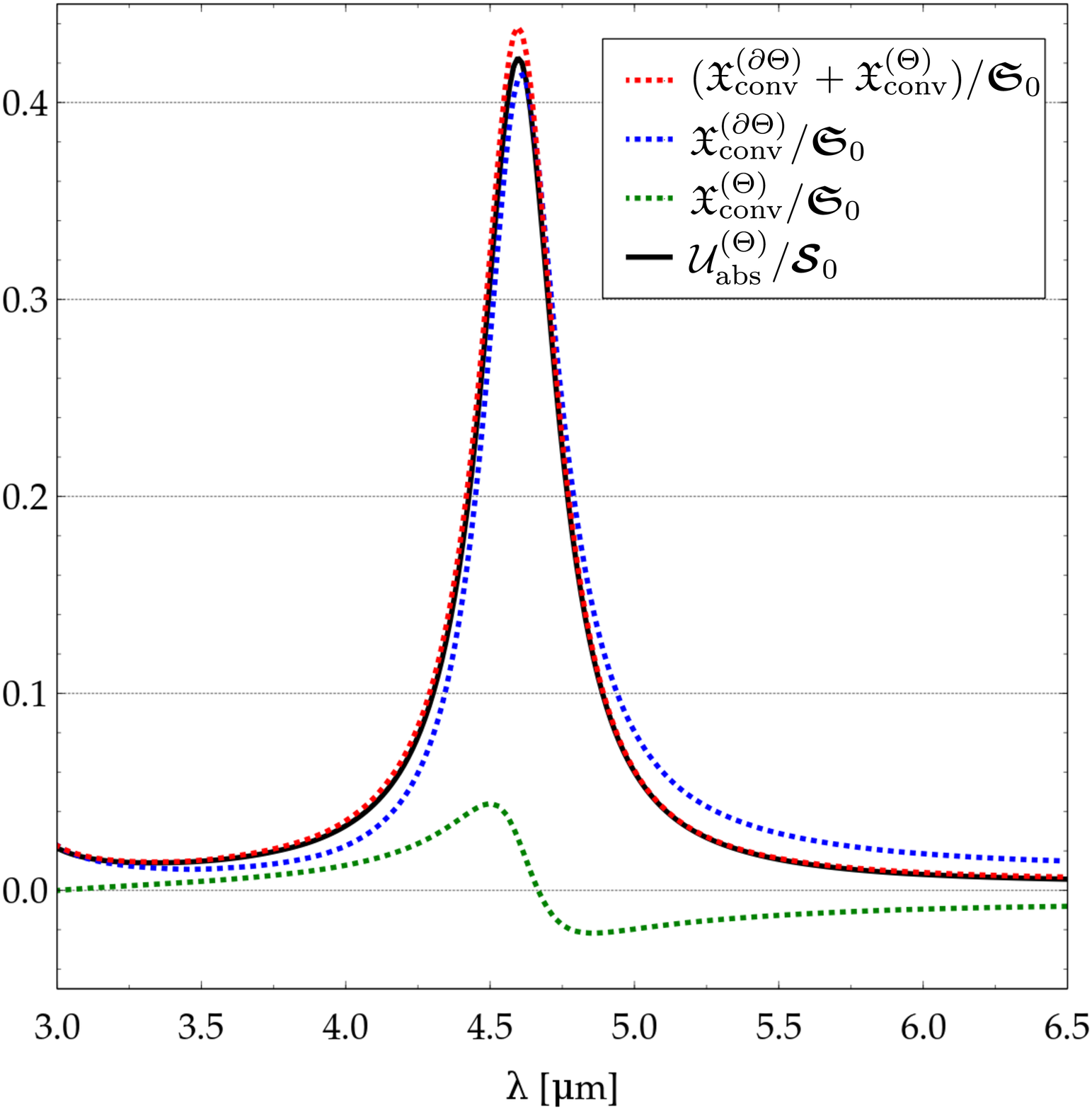}
\end{tabular}
}
\\
(a) setup &
(b) transmission for CPL &
(c) chiral vs.\ energy quantities for RPL
\end{tabular}
\end{center}
\caption{
	\label{fig:helixSetup}
	Gold helix surrounded by air in hexagonal lattice with lattice constant $\sqrt{3} a$ (a). The helix is illuminated by a plane wave
	from above.
	The geometrical parameters $h, r$ and $d$ are chosen in such a way that the helix is sharply resonant for RPL
	and very close to transparent for LPL.
	This is visible in the transmission spectrum (b, top) and the conversion from RPL to LPL and vice versa (b, bottom).
	RPL (blue) exhibits a transmission dip while LPL (red) is nearly fully transmitted over the given wavelength range.
	Note the scaling of the $y$-axis
	in the conversion plot: there is nearly no conversion from one circular polarization to another.
	In the spectra for RPL (c, top) 
	the transmitted (black) and reflected (grey) energy flux $\myInt{\Sth}{\dd{\Omega}}$ are proportional to
	the transmitted (dashed red) and reflected (dashed blue) chirality flux $\myInt{\Sigmth}{\dd{\Omega}}$.
	The simple relation of energy and chiral quantities is due to the low polarization conversion (b).
	This is also reflected in the comparison of energy absorption and chirality conversion (c, bottom).
	The shown graphs correspond to:
	absorbed energy (black), converted chirality in the volume $\myInt{\chith_\text{conv}}{\Theta}$ (dashed green),
	converted chirality at the interface $\myInt{\chith_\text{conv}}{\dd{\Theta}}$ (dashed blue)
	and the total converted chirality as sum of the two latter (dashed red).
	Again, the proportionality of the summed converted chirality and the absorbed energy is clearly visible.
	The presented formalism introduces the decomposition of conversion of chirality into surface
	$\myInt{\chith_\text{conv}}{\dd{\Theta}}$ and volume $\myInt{\chith_\text{conv}}{\Theta}$ terms.
}
\end{figure}

In order to establish the conservation of optical chirality \eqref{eq:chConsth} as a practical numerical tool,
we analyse an array of gold helices in a design similar to the experiment of Gansel et al.\ \cite{gansel2009} in the infra-red wavelength range.
The general setup and our geometry parameters are shown in Fig.\ \ref{fig:helixSetup} (a). The gold helix is surrounded by air ($n=1$)
and placed in a hexagonal unit cell which yields a slab of the array of helices. The used refractive index of gold is given by the literature \cite{palik1998}.
The height of the helix is $h=2\mu\text{m}$, the wire diameter equals $d=0.2\mu\text{m}$ and the radius is chosen as $r=0.2\mu\text{m}$. 
The lattice constant is $\sqrt{3}a = 2\mu\text{m}$.
The substrate is omitted.

For this as well as for all simulations of this study, we use the FEM solver \textit{JCMsuite} \cite{pomplun2007,JCMsuite} wherein the chirality densities for
isotropic media (\ref{eq:chielJCM}, \ref{eq:chimaJCM}) as well as for anisotropic media (\ref{eq:chielth}, \ref{eq:chimath}) and
the chirality flux density \eqref{eq:sigmth} are readily implemented and in which we use finely tuned numerical parameters
to ensure well converged results.
For the stated results, we use locally adapted polynomial degrees $p$ of the finite elements,
so-called $hp$-FEM \cite{burger2015},
to both model the spatial discretization of the helix fine enough and reduce computational resources.
Using this approach, we chose the numerical parameters such that violations of conservation of chirality \eqref{eq:chConsth} due to numerical
discretization errors are of less than $1 \%$ for each point of a wide wavelength scan.

\begin{figure} [b]
\begin{center}
\begin{tabular}{cc} 
\includegraphics[height=6cm]{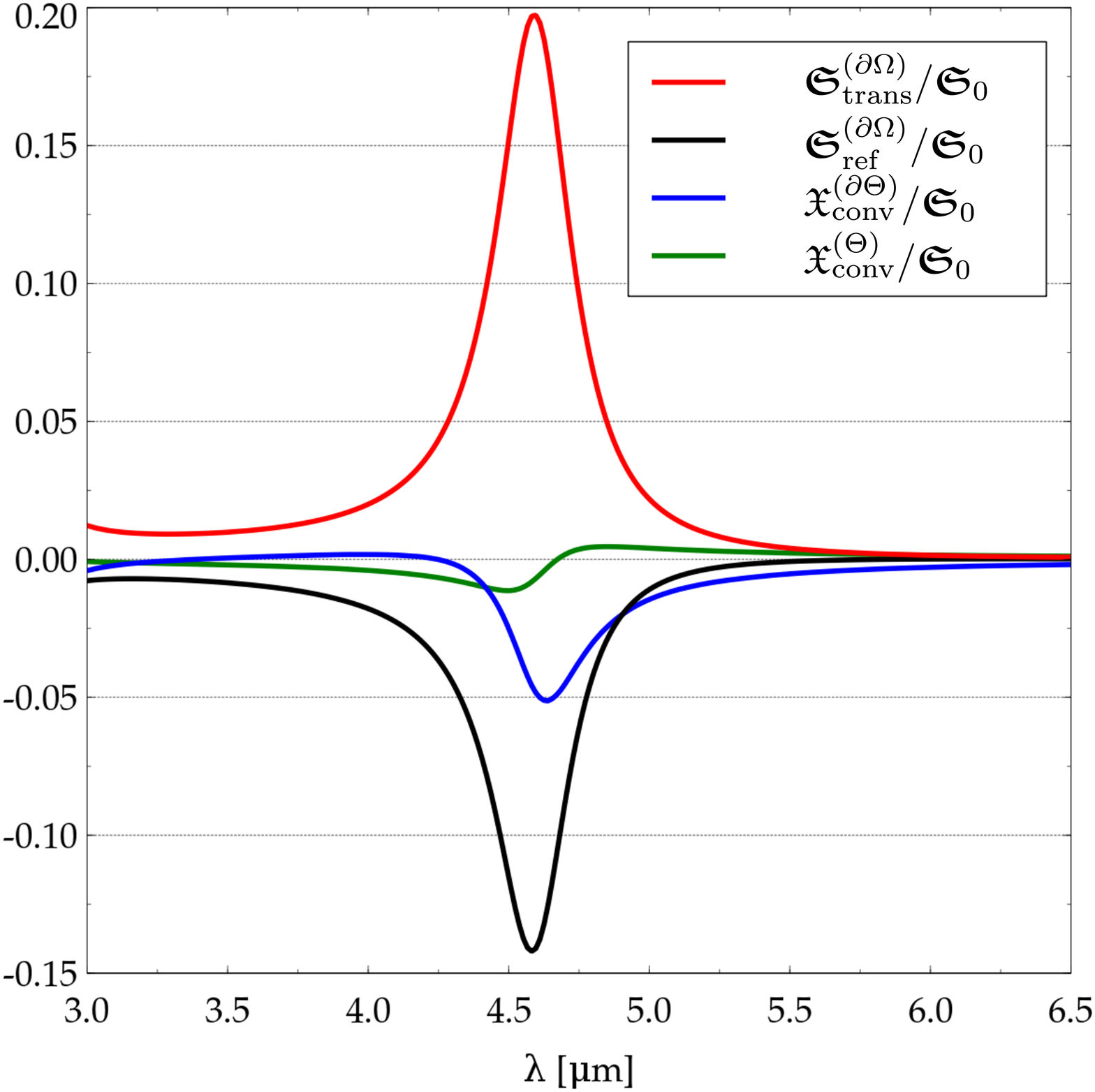} &
\hspace{1cm}
\includegraphics[height=6cm]{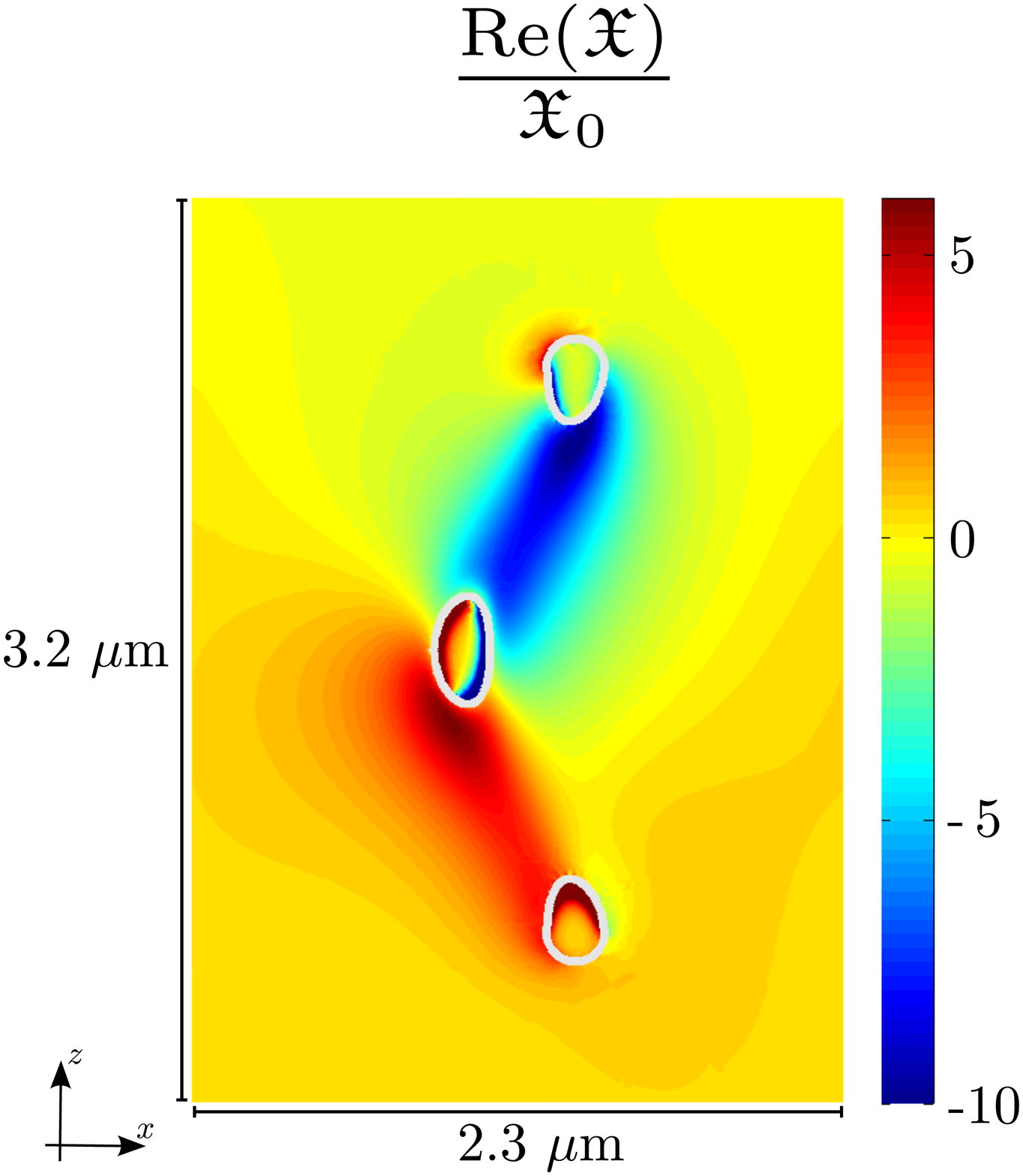}
\\
(a) chirality conservation &
\hspace{1cm}
(b) near-field chirality
\end{tabular}
\end{center}
\caption{
	\label{fig:chirality_x}
	Conservation of optical chirality in the helix array of Fig.\ \ref{fig:helixSetup} for illumination with $x$-polarized light (a).
	Since the incoming chirality flux of the linearly polarized plane wave is zero, conservation of chirality
	predicts that the sum of the transmitted (black), reflected (red) chirality flux,
	chirality conversion on the interface (green) and chirality conversion in the volume (blue) is zero as well.
	There exists a distinct change of sign for the converted chirality in the volume $\myInt{\chith_\text{conv}}{\Theta}$
	similar to RPL illumination [cf.\ Fig.\ \ref{fig:helixSetup} (c, bottom)].
	In (b), we show the near-field chirality in the $xz$-plane for the same linearly polarized illumination and $\lambda_0 = 4.6\mu\text{m}$.
	The point-wise or averaged optical chirality in the near-field is expected to enhance CD measurements of 
	chiral molecules \cite{schaeferling2014,tian2015,poulikakos2015}.
}
\end{figure}

We optimize the device to be sharply resonant only for one circular polarization, namely right polarized light
[Fig.\ \ref{fig:helixSetup} (b)].
Note that there is nearly no conversion from RPL to LPL and vice versa in transmission direction.
Accordingly, this device is an approximation of an electromagnetically chiral object which would be completely transparent to one circular polarization \cite{fernandez2015}.
Since there is only one diffraction order in this setup, the conversion can, on the one hand, be computed conventionally by projecting the obtained
Fourier coefficients onto the right and left circular polarization basis. Nevertheless, this requires a specified propagation direction
of the far-field propagating plane waves.

On the other hand, one can use the decomposition of the electromagnetic energy flux into right and left circularly polarized components with the
help of chirality flux analogous to the decomposition of energy (\ref{eq:chEn}, \ref{eq:chEnEn}).
This near-field decomposition does not require any propagation direction and yields equal results to the Fourier polarization basis
for this periodic setup with one diffraction
order. Furthermore, in contrast to the simple Fourier basis of circular polarization this concept is generally valid for non-periodic structures
as well as for arbitrary number of diffraction orders.

We use the given setup to illustrate the proportionality of
energy and chirality \cite{bliokh2011}, however, our findings are also valid for the general case with arbitrary conversion between
incident, reflected and transmitted polarizations.
Since there is low polarization conversion, transmitted and reflected chirality flux and electromagnetic
energy flux are proportional for e.g.\ RPL which is shown in Fig.\ \ref{fig:helixSetup} (c, top).

Although the energy is purely absorbed within the helix ($\Theta$), the optical chirality is converted at both the interface ($\dd{\Theta}$)
via the gradient of the real part of permittivity as well as within the helix through non-vanishing imaginary part of the permittivity.
With our formalism, we are able to attribute the conversion of chirality separately to the shape and the volume
of the scatterer [Fig.\ \ref{fig:helixSetup} (c, bottom)].
To the best of our knowledge, this has not been formulated in the literature before and further deepens the
understanding of the quantity of optical chirality.
This point of view might be applicable for further analysis of the angular momenta of light, namely orbital and spin
angular momenta \cite{bliokh2015transverse}.

Again, since our numerical example exhibits nearly no conversion from RPL to LPL,
the total conversion of chirality consisting of surface and volume contributions is proportional to the absorbed energy.
Small deviations are due the small conversion from RPL to LPL which has its maximum at the resonance dip of
transmittance [Fig.\ \ref{fig:helixSetup} (b)].

With the herein given conservation law of time-harmonic optical chirality,
we can also analyse linearly polarized light and its chiral quantities (Fig.\ \ref{fig:chirality_x}).
Most importantly the picture of this conservation law is implicitly generalized for
every setup with arbitrary material parameters and incident illuminations. It enables us to compare chiral far-field
[Fig.\ \ref{fig:chirality_x} (a)] as well as near-field [Fig.\ \ref{fig:chirality_x} (b)] quantities:
the resonances of the chirality conversions in the near-field are slightly red-shifted with respect to the far-field
chirality flux resonances. Furthermore, linearly polarized light generates homogeneous optical chirality of opposite
handedness in the lower and the upper part of the helix, respectively.  

Especially, the proportionality between chirality flux and energy flux for this helix sample is not always
given.
The conservation law, however,
is valid in all circumstances, i.e.\ also for structures with high conversions between circular polarization
states, for incident linear polarization as well as non-periodic structures in which the well-known concept
of extinction in the case of energy conservation can be translated to chiral extinction \cite{nieto2015}
straightforwardly.

\section{CONCLUSION}
\label{sec:conc}

We study a novel conservation law of optical chirality in inhomogeneous isotropic \eqref{eq:chConsth} and arbitrary space \eqref{eq:chContth}
\cite{poulikakos2015}.
This has been achieved by first deriving the well-known point-wise continuity equation \cite{tang2010} for arbitrary media
including anisotropic material parameters yielding a converted chirality density in analogy to absorbed energy.
Second, we identified loss, anisotropy and gradients in the real part of the material parameters permittivity and permeability
as generators of conversion of chirality.

We expect that chirality conversion of piecewise homogeneous media
provides insights to the influence of geometry in order to optimize structures for high local chirality density and applications such as CD
spectroscopy \cite{mcPeak2014}. 
Furthermore, the chirality conversion on interfaces might be illustrative for the analysis of angular momenta of light and its
decomposition into spin and orbital parts \cite{bliokh2015transverse}.
The given conservation law of optical chirality
establishes an additional tool for the analysis of these effects within numerical simulations such as FEM simulations. 

In a numerical example, we showed the close connection of chirality and energy \cite{bliokh2011}.
Furthermore, our formalism provides insight in the conversion of chirality by discriminating surface and volume terms which
is impossible without chiral and with solely conventional electromagnetic energy quantities.
Additionally, the presented theory is also valid for an arbitrary number of diffraction orders and non-periodic setups.
Due to its generality, the formalism can be extended in analogy to extinction energy to the far-field
effects of scattered and converted chirality.

This paves the way for the analysis and rigorous definition of chiral analogues to the Purcell factor and the mode volume \cite{yoo2015} as well
as scattering and absorption cross-sections \cite{nieto2015}.

\pagebreak

\appendix

\section{CHIRALITIY CONTINUITY EQUATION}
\label{app:cont}

For the derivation of the chirality continuity equation \eqref{eq:chContth}, we use the vector identity
\begin{align}
    \nabl \cdot (\Eth \times \Hth) = \Hth \cdot (\nabl \times \Eth) - \Eth \cdot (\nabl \times \Hth)
    \label{eq:vecId}
\end{align}
and equivalent identities for the rotation of the fields
\begin{align}
	\nabl \cdot [(\nabl \times \Eth) \times \Hth] &= \Hth \cdot [\nabl \times (\nabl \times \Eth)] - (\nabl \times \Eth) \cdot (\nabl \times \Hth)
	\label{eq:vecId1} \\
	\nabl \cdot [\Eth \times (\nabl \times \Hth)] &=  (\nabl \times \Hth) \cdot (\nabl \times \Eth) - \Eth \cdot [\nabl \times  (\nabl \times \Hth)].
	\label{eq:vecId2}
\end{align}
Using the time-harmonic analogue of the suggested time-dependent quantity \cite{tang2010} to derive the continuity equation of optical chirality,
we obtain
\begin{align*}
	&\underline{\Jth^*} \cdot \rot{\Eth} + \Eth \cdot \rot{\underline{\Jth^*}} 
	\\	
	&\stackrel{\eqref{eq:Maxwell2th}}{=} {\color{red}\underline{\rot{\Hth^*}} \cdot \rot{\Eth}} - \underline{i \omega \Dth^*} \cdot \rot{\Eth}
		+ \Eth \cdot [\nabl \times \underline{\rot{\Hth^*}}]
		- \underline{i \omega} \Eth \cdot \underline{\rot{\D^*}}
	\\
	&\stackrel{{\color{red}\eqref{eq:vecId1}}}{=} {\color{red}\Hth^* \cdot [\nabl \times \rot{\Eth}]} - i \omega \Dth^* \cdot \rot{\Eth}
		+ \underline{\Eth \cdot [\nabl \times \rot{\Hth^*}]}
		- i \omega \Eth \cdot \rot{\D^*}
			{\color{red} - \nabl \cdot \left[ \rot{\Eth} \times \Hth^* \right]}
	\\	
	&\stackrel{\eqref{eq:vecId2}}{=} \Hth^* \cdot [\nabl \times \rot{\Eth}] - i \omega \Dth^* \cdot \rot{\Eth}
		+ \underline{\rot{\Hth^*} \cdot \rot{\Eth}}
		- i \omega \Eth \cdot \rot{\D^*} \nonumber
		\\
		&\hspace{1cm}
			{\color{red}- \nabl \cdot \left[ \rot{\Eth} \times \Hth^* \right] \underline{- \nabl \cdot \left[ \Eth \times \rot{\Hth^*} \right]}}
	\\
	&\stackrel{{\color{red}s.r.}}{=} \Hth^* \cdot [\nabl \times \underline{\rot{\Eth}}] - i \omega \Dth^* \cdot \rot{\Eth}
		+ \rot{\Hth^*} \cdot \underline{\rot{\Eth}}
		- i \omega \Eth \cdot \rot{\D^*} \nonumber
		\\
		&\hspace{1cm}
			{\color{red}- \nabl \cdot \left[ \Eth \times \rot{\Hth^*} - \Hth^* \times \rot{\Eth} \right]} \label{eq:contthStep}
	\\
	&\stackrel{\eqref{eq:Maxwell1th}}{=} \underline{i \omega} {\color{red}\Hth^* \cdot [\nabl \times \underline{\Bth}]} - i \omega {\color{red}\Dth^* \cdot \rot{\Eth}}
		+ \underline{i \omega} {\color{red}\rot{\Hth^*} \cdot \underline{\Bth}}
		- i \omega {\color{red}\Eth \cdot \rot{\D^*}} \nonumber
		\\
		&\hspace{1cm}
			- \nabl \cdot \left[ {\color{red}\Eth \times \rot{\Hth^*} - \Hth^* \times \rot{\Eth}} \right]
	\\
	&\stackrel{{\color{red}(\ref{eq:chielth}-\ref{eq:sigmth})}}{=} {\color{red}8} i \omega {\color{red}\chimath}
		- {\color{red}8} i \omega {\color{red}\chielth} - \nabl \cdot {\color{red}4 \Sigmth},
\end{align*}
where $s.r.$ is the common sum rule of derivatives and the terms involved in each step are marked in red or underlined, respectively.
Scaling similar to the energy continuity equation yields \eqref{eq:chContth}
\begin{align*}
	2 i \omega (\chielth - \chimath) + \nabl \cdot \Sigmth = -\frac{1}{4} \left[ \Jth^* \cdot \rot{\Eth} + \Eth \cdot \rot{\Jth^*} \right].
\end{align*}

\section{UPPER BOUND OF CHIRALITY}
\label{app:bound}

\newcommand{\e}{\vec{e}}
\newcommand{\h}{\vec{h}}

Here, we restate the upper bound of chirality \cite{bliokh2011} within our formalism to obtain \eqref{eq:chEn} for
isotropic lossless media, i.e.\ $\epsi,\mue \in \mathbb{R}^{+}$.
We start with Young's inequality for the special case of an electric $\e \in \{\imag{\Eth}, \real{\Eth}\}$ and a magnetic
$\h \in \{\imag{\Hth}, \real{\Hth}\}$ field, where the impedance $Z = \sqrt{\mue/\epsi}$ is required:
\begin{align}
	0 &\leq \left( \e - \sqrt{\frac{\mue}{\epsi}} \h \right)^2
		= \e^2  + \frac{\mue}{\epsi} \h^2 - 2 \sqrt{\frac{\mue}{\epsi}} \e \cdot \h \\
	\e \cdot \h &\leq \frac{1}{2} \left( \sqrt{\frac{\epsi}{\mue}} \e^2 + \sqrt{\frac{\mue}{\epsi}} \h^2 \right).
	\label{eq:young}
\end{align}
Furthermore, we use
\begin{align}
	\sqrt{\epsi \mue} = \frac{n}{c_0}.
	\label{eq:refInd}
\end{align}
Starting from \eqref{eq:chIsoLossless}, we obtain \eqref{eq:chEn}:
\begin{align*}
	\left| \chith \right| &= \left| \frac{1}{2} \omega \imag{\Dth^* \cdot \Bth} \right|
		= \frac{1}{2} \omega \left| \epsi \mue \imag{\Eth^* \cdot \Hth} \right| 
		= \frac{1}{2} \omega \left| \epsi \mue
			\left[ \real{\Eth} \cdot \imag{\Hth} - \imag{\Eth} \cdot \real{\Hth} \right] \right| \\
		&\leq \frac{1}{2} \omega \left| \epsi \mue \right| \left[ \left| \real{\Eth} \cdot \imag{\Hth} \right|
			+ \left| \imag{\Eth} \cdot \real{\Hth} \right| \right] \\
		&\stackrel{\eqref{eq:young}}{\leq}
			\frac{1}{4} \omega \left| \epsi \mue \right| \left[
				\left| \sqrt{\frac{\epsi}{\mue}} \real{\Eth}^2 + \sqrt{\frac{\mue}{\epsi}} \imag{\Hth}^2 \right|
				+  \left| \sqrt{\frac{\epsi}{\mue}} \imag{\Eth}^2 + \sqrt{\frac{\mue}{\epsi}} \real{\Hth}^2 \right|
				\right]	\\
		&\stackrel{\eqref{eq:refInd}}{\leq} \frac{1}{4} \omega \left| \frac{n}{c_0} \sqrt{\epsi \mue} \right|
				\left[
				\left| \sqrt{\frac{\epsi}{\mue}} \Eth \right|^2 + \left| \sqrt{\frac{\mue}{\epsi}} \Hth \right|^2 
				\right]
		= \frac{1}{4} \frac{\omega n}{c_0} \left[
				\epsi \left| \Eth \right|^2 +
				\mue \left| \Hth \right|^2 \right] \\
		&\leq \frac{\omega n}{c_0} \left[ \Uelth + \Umath \right]
		 = \frac{\omega n}{c_0} \Uth.
\end{align*}

\section{CHIRALITY CONVERSION ON INTERFACES}
\label{app:convInt}

\newcommand{\ee}{\vec{\mathfrak{E}}}

In order to derive the influence of spatial varying isotropic $\epsi \in \mathbb{C}$, we use
\begin{align}
	\nabl \times \left(\epsi^* \Eth\right) = \epsi \rot{\Eth} + \left(\nabl\epsi^*\right) \times \Eth.
	\label{eq:grad}
\end{align}
Furthermore, we use the invariance under a circular shift of the triple product
\begin{align}
	\label{eq:triple}
	\Eth \cdot \left[ \left(\nabl \epsi^*\right) \times \Eth^* \right] = \left(\nabl \epsi^*\right) \cdot \left(\Eth^* \times \Eth \right),
\end{align}
as well as the fact that
\begin{align}
	\Eth^* \times \Eth = i \imag{\Eth^* \times \Eth}.
	\label{eq:EtE}
\end{align}
For the electric optical chirality, we obtain from \eqref{eq:chielth}
\begin{align*}
	8 \chielth &=
		\Dth^* \cdot \rot{\Eth} + \Eth \cdot \rot{\Dth^*} \\
		&\stackrel{\eqref{eq:grad}}{=}
			\epsi^* \Eth^* \cdot \rot{\Eth} + \epsi^* \Eth \cdot \rot{\Eth^*} + \Eth \cdot \left[ \left(\nabl \epsi^*\right) \times \Eth^* \right] \\
		&\stackrel{\eqref{eq:triple}}{=}
			2 \epsi^* \operatorname{Re}\left[ \Eth^* \cdot \rot{\Eth} \right]
			+ \left(\nabl \epsi^*\right) \cdot \left( \Eth^* \times \Eth \right) \\
		&\stackrel{\eqref{eq:EtE}}{=}
			2 \epsi^* \operatorname{Re}\left[ \Eth^* \cdot \rot{\Eth} \right]
			+ i \left(\nabl \epsi^*\right) \cdot \imag{\Eth^* \times \Eth} \\
		&\stackrel{\eqref{eq:Maxwell1th}}{=}
			2 \omega \epsi^* \real{i \Eth^* \cdot \Bth}
			+ i \left(\nabl \epsi^*\right) \cdot \imag{\Eth^* \times \Eth} \\
		&=
			-2 \omega \epsi^* \imag{ \Eth^* \cdot \Bth }
			+ i \left(\nabl \epsi^*\right) \cdot \imag{\Eth^* \times \Eth}
\end{align*}
and, accordingly, for the conversion of chirality at interfaces \eqref{eq:chConvInt}
\begin{align*}
	8 \imag{\chielth} &=
			2 \omega \imag{\epsi} \imag{ \Eth^* \cdot \Bth }
			+ \left[ \nabl \real{\epsi} \right] \cdot \imag{\Eth^* \times \Eth}
\end{align*}

\acknowledgments
We thank Lin Zschiedrich for many fruitful discussions.  
We acknowledge the support of BMBF through project 13N13164
(SolarNano) and
of the Einstein Foundation Berlin through projects ECMath-OT5 and -SE6,
as well as support from the Deutsche Forschungsgemeinschaft (DFG) through SFB
787 'Semiconductor Nanophotonics: Materials, Models, Devices' and 
from the Zuse Institute Berlin through bridge project
'Parameter-dependent parallel block sparse Arnoldi and D\"ohler algorithms on distributed systems'.

\bibliography{references} 
\bibliographystyle{spiebib} 

\end{document}